\documentclass{emulateapj}
\usepackage{graphicx}
\usepackage{bm}
\usepackage{amssymb}
\usepackage{amsmath}
\usepackage{cancel}
\usepackage{hyperref}
\usepackage{physics}

\usepackage{savesym}
\savesymbol{tablenum}
\usepackage{siunitx}
\restoresymbol{SIX}{tablenum}
\DeclareSIUnit\mpc{Mpc}
\DeclareSIUnit\solarmass{M_{\sun}}
\DeclareSIUnit\solarluminosity{L_{\sun}}
\DeclareSIUnit\year{yr}

\usepackage{multirow}

\newcommand{\SFR}{\text{SFR}}

\newcommand{\CII}{\text{CII}}

\newcommand{\Var}{\mathrm{Var}}
\newcommand{\Cov}{\mathrm{Cov}}

\usepackage{epstopdf}
\usepackage{natbib}
\usepackage{epsfig}
\usepackage{cleveref}

 \newcommand{\lya}{Ly$\,\alpha$ }

\newcommand{\avg}[1]{\ensuremath{\langle #1 \rangle}}
\newcommand{\bma}{\begin{math}}
\newcommand{\ema}{\end{math}}
\newcommand{\beq}{\begin{equation}}
\newcommand{\eeq}{\end{equation}}
\newcommand{\beqa}{\begin{eqnarray}}
\newcommand{\eeqa}{\end{eqnarray}}
\newcommand{\bc}{\begin{center}}
\newcommand{\ec}{\end{center}} 
\newcommand{\bit}{\begin{itemize}}
\newcommand{\eit}{\end{itemize}}

\font\BFd=cmmib10
\font\BFt=cmmib10
\font\BFs=cmmib10 scaled 700
\font\BFss=cmmib10 scaled 500

\def\bbox#1{%
\relax\ifmmode
\mathchoice
{{\hbox{\BFd #1}}}
{{\hbox{\BFt #1}}}
{{\hbox{\BFs #1}}}
{{\hbox{\BFss #1}}}
\else \mbox{#1} \fi }

\def\k{{\bbox{k}}}
\def\q{{\bbox{q}}}

\def\x{{\bbox{x}}}

\newcommand{\tmin}{\text{min}}
\newcommand{\tmax}{\text{max}}
\newcommand{\Mpc}{\text{Mpc}}
\newcommand{\Gpc}{\text{Gpc}}

\newcommand{\HI}{\text{HI}}
\newcommand{\mK}{\text{mK}}
\newcommand{\Jystr}{\text{Jy}/\text{str}}

\newcommand{\bto}{\avg{T_{21}}b_{21}}

\begin{document}

\title{Extracting bias using the cross-bispectrum: An EoR and 21~cm-[CII]-[CII] case study}
\author{Angus Beane\altaffilmark{1} \& Adam Lidz\altaffilmark{1}}
\altaffiltext{1} {Department of Physics \& Astronomy, University of Pennsylvania, 209 South 33rd Street, Philadelphia, PA 19104, USA}

\email{abeane@sas.upenn.edu}

\begin{abstract}
The amplitude of redshifted 21~cm fluctuations during the Epoch of Reionization (EoR) is
expected to show a distinctive ``rise and fall'' behavior with decreasing redshift as reionization proceeds. On large scales ($k \lesssim 0.1$ Mpc$^{-1}$) this can mostly be
characterized by evolution in the product of the mean 21~cm brightness temperature and a bias factor, $\bto(z)$.
This quantity evolves in a distinctive way that can help in determining the average ionization history of the intergalactic medium (IGM) from upcoming 21~cm fluctuation data sets.
Here we consider extracting $\bto(z)$ using a combination of future redshifted 21~cm and [CII] line-intensity mapping data sets. Our method exploits the dependence of the 21~cm-[CII]-[CII]
cross-bispectrum on the shape of triangle configurations in Fourier space. This allows one to determine $\bto(z)$ yet, importantly, is less sensitive to foreground contamination than
 the 21~cm auto-spectrum, and so can provide a valuable cross-check. 
We compare the results of simulated bispectra with second-order perturbation theory: on the largest scales well-probed by our simulations ($k \sim 0.05\,\Mpc^{-1}$), the perturbative estimate
of $\bto$ matches the true value to within $10\%$ for $\avg{x_i} \lesssim 0.8$. 
The perturbative formula is most accurate early in the EoR.
We consider the 21 cm auto-bispectrum and show that this statistic may also be used to extract the 21~cm bias factor. Finally, we discuss the survey requirements for measuring the cross-bispectrum. Although we focus on the 21~cm-[CII]-[CII] bispectrum during reionization, our method may be of broader interest and can be applied to any two fields throughout cosmic history.
\end{abstract}


\section{Introduction} \label{sec:intro}

Observations of the redshifted 21~cm line promise to reveal the timing and spatial structure of the Epoch of Reionization (EoR) in the near future. This will help determine the formation time and properties of the first stars, galaxies, and accreting black holes, and the nature of large-scale structure at high redshift \citep{Loeb13}. In order to understand the full astrophysical implications of the upcoming data, a range of challenges must first be overcome, both to mitigate systematic effects from foreground contamination and instrumental artifacts, and to faithfully model and interpret the measurements. 

One goal of redshifted 21~cm surveys is to measure the redshift evolution of the power spectrum of 21~cm fluctuations. On large scales, the amplitude of the 21~cm power spectrum is expected to rise and fall with decreasing redshift as reionization proceeds (e.g. \citealt{Lidz08}). This can in turn be used to extract information about the volume-averaged ionization fraction and its redshift evolution. On scales larger than the size of the ionized bubbles, this redshift evolution should mostly be characterized by a bias factor, $b_{21}(z)$, relating the 21~cm fluctuations to fluctuations in the underlying matter density field on large scales \citep{furlanetto04:hiigrowth}, and by the spatial average 21~cm brightness temperature $\avg{T_{21}}(z)$.
The bias factor description, if accurate, has the important virtue of providing a model-independent characterization of the 21~cm fluctuation measurements. The measured bias factors can then be compared with simulations to extract information about the ionizing sources and the reionization history.

In conjunction with the 21~cm experiments, a number of efforts are underway to carry out line-intensity mapping surveys of the EoR in various other emission lines, such as [CII] \citep{Crites14}, CO \citep{Chung:2017uot}, and \lya \citep{Dore16}. In these line-intensity mapping observations, one measures the spatial fluctuations in the collective emission from many individually unresolved sources \citep{kovetz2017:im_review}. Like the 21~cm experiments, these observations span large regions on the sky and provide accurate redshift information, enabling cross-correlation measurements with the 21~cm data sets \citep{Lidz11,gong12:during,gong11:probing,silva15:prospects}. 
Surveys in lines such as [CII] and CO should trace large-scale structure in the galaxy distribution and complement the redshifted 21~cm experiments, which probe the IGM \citep{Lidz:2008ry}. 
(\lya provides a useful intermediate case, reflecting fluctuations in both the source distribution and that of neutral hydrogen in the IGM, e.g. \citealt{Pullen:2013dir}.) In addition, the line-intensity 21~cm cross-correlation is less susceptible to foreground contamination and other systematic effects than the auto-spectrum extracted from either data set alone \citep{Furlanetto:2006pg,Lidz:2008ry,Lidz11}. The advantage of the cross-spectrum is that residual 21~cm foregrounds from 
galactic synchrotron, for example, may produce a strong bias in the auto-spectrum yet do not correlate on average with other tracers of high redshift structure.

Here we consider an approach for extracting $\bto(z)$ from a combination of line-intensity mapping and redshifted 21~cm data. Our aim is to develop a way of extracting this key quantity from upcoming data that is less susceptible to systematic contamination than the usual 21~cm auto-spectrum measurements. In this work we focus on the [CII] line, since surveys are already underway to detect this line from the EoR \citep{Crites14,Lagache:2018hmk}, but related measurements could be carried out using other lines.

The method developed here exploits the fact that the growth of large-scale structure under gravity generates non-Gaussianity, and that the resulting matter bispectrum has a distinctive dependence
on triangle configuration. In the context of galaxy surveys, this has long been recognized and used to constrain galaxy bias (e.g. \citealt{Fry94,Matarrese:1997sk,Scoccimarro:2000sp,Verde:2001sf}). 
In principle, a closely related measurement may be used to constrain 21~cm biasing as recently discussed by \citet{Hoffman2018:2point3point} in the context of the three-point correlation of the
21~cm field in configuration space. This recent work follows a number of papers on the general theme of  using the 21~cm bispectrum to complement 21~cm power spectrum measurements during reionization \citep{Bharadwaj:2004sx,Shimabukuro:2015iqa,Shimabukuro:2016viy,Majumdar:2017tdm}. In addition to their utility in constraining 21~cm biasing, higher-order statistics are required to access the full information content of the highly non-Gaussian 21~cm signal expected during reionization. 

Here we consider a variant of the usual approach for constraining biasing: we propose to extract $\bto(z)$ from a cross-bispectrum statistic, specifically the 21~cm-[CII]-[CII] bispectrum (involving one 21~cm field and two [CII] fields).
This statistic has the virtue that it is less sensitive to foreground contamination than the 21~cm auto-spectrum and the 21~cm auto-bispectrum (for similar reasons to the two-point cross-correlation) , and can potentially provide a powerful cross-check of inferences from the 21~cm auto-spectrum. We also consider extracting $\bto(z)$ from the 21~cm auto-bispectrum, and we find that this provides another route for constraining  21~cm biasing, in broad agreement with earlier work from \citet{Hoffman2018:2point3point}. In contrast to this previous study, we work in Fourier space since this is a more natural basis for the interferometric 21~cm measurements.
The 21~cm auto-bispectrum is, however, more sensitive to foreground contamination than the cross-bispectrum advocated here. On the other hand, it is likely a better probe of reionization's early phases since [CII] emission may be dim at early times, i.e. before metal enrichment is well underway.

The outline of this paper is as follows. In \S~\ref{sec:approach} we present our approach for measuring the redshift evolution of the 21~cm bias factor. In \S~\ref{sec:methods} we discuss the reionization simulations used to develop and test our analysis technique 
and the algorithm used to measure the bispectrum from simulations. We then discuss the simulated cross-bispectrum measurements in \S~\ref{sec:results}, and consistency tests of the perturbative framework in \S~\ref{sec:consistency_checks}. In \S~\ref{sec:detectability} we discuss the prospects for measuring the cross-bispectra from future surveys. We summarize our results in \S~\ref{sec:conclusion}. We assume a $\Lambda$CDM cosmology, parameterized by
$(\Omega_m, \Omega_{\Lambda}, \Omega_b, h, \sigma_8, n_s) = (0.27, 0.73, 0.046, 0.7, 0.8, 1)$, in broad consistency with recent Planck measurements \citep{Ade:2015planck}

\section{Constraining 21~cm Bias with the Cross-Bispectrum}\label{sec:approach}

Here we explain our technique, describing how the 21~cm bias may be extracted from the 21~cm-[CII]-[CII] cross-bispectrum. Consider first the 21~cm brightness temperature field, which describes
the brightness temperature contrast between a neutral hydrogen cloud and the cosmic microwave background:
\beq
\label{eq:t21}
T_{21}(\x) = T_0 X_{\HI}(\x) \left[1 + \delta_\rho(\x)\right],
\eeq
Here $T_0 = 28.1745\,\text{mK} \left[(1+z)/10\right]^{1/2}$ (e.g. \citealt{Zaldarriaga:2003du}), $X_{\HI}(\x)$ is the neutral hydrogen fraction (at position $\x$), and $\delta_\rho(\x)$ is the gas density contrast, which is assumed
to follow the overall matter density field on the large scales of interest. We can further define the fractional 21~cm brightness temperature fluctuation, 
$\delta_{21}(\x) = (T_{21}(\x) - \avg{T_{21}})/\avg{T_{21}}$. Since $\avg{T_{21}}$ is not directly observable in an interferometric measurement, we will
ultimately work with the statistics of $T_{21}(\x)$, or more precisely its Fourier counterpart, $T_{21}(\k)$.\footnote{Our Fourier convention is: $T_{21}(\k) = \int \text{d}^3x\, T_{21}(\x) e^{i \k \cdot \x}$ and $T_{21}(\x) = \int \frac{\text{d}^3k}{(2\pi)^3}\, T_{21}(\k) e^{-i \k \cdot \x}$.}
 However, it is cleaner to work with dimensionless fluctuation fields, such as $\delta_{21}(\x)$, in
describing biasing relations. 
Eq.~\ref{eq:t21} ignores the impact of peculiar velocities and spin temperature fluctuations, but
these should be good approximations during most of the EoR (e.g. \citealt{Jensen:2013fha,Pritchard:2006sq}).

Similarly, we can consider the spatial fluctuations in the [CII] emission field, as probed by
upcoming [CII] line-intensity mapping experiments. The [CII] emission is generally characterized by the three-dimensional specific intensity field, denoted here
by $I_{\CII}(\x)$ \citep{Lidz11,kovetz2017:im_review}, or its Fourier partner, $I_{\CII}(\k)$. The fractional fluctuations in the [CII] specific intensity field are $\delta_{\CII}(\x) = (I_{\CII}(\x) - \avg{I_{\CII}})/\avg{I_{\CII}}$. 

As we will test subsequently using simulations of the EoR, we assume that on large scales each of these fluctuation fields is a deterministic and local function of the underlying matter density 
distribution \citep{Scherrer:1997hp,Dekel:1998eq}.
For example, we suppose that $\delta_{21}(\x) = f \left[\delta_\rho(\x)\right]$, where $f$ is the biasing function that specifies the relation between the 21~cm and
matter density fluctuations. It is not obvious that the local biasing assumption is adequate for describing 21~cm
fluctuations during reionization, since distant sources may impact the ionization state
of the gas -- especially at the late stages of reionization. However, as we will quantify, 
local biasing may nevertheless be a
good approximation if we consider sufficiently large scales. Considering such scales, we Taylor expand the biasing function for small $\delta_\rho(\x)$, keeping terms to second-order in the density fluctuations:
\beq
\label{eq:bias_expand}
\delta_{21}(\x) = b_{21} \delta_\rho(\x) + \frac{b_{21}^{(2)}}{2} \delta^2_{\rho}(\x).
\eeq 
This equation defines the first-order and second-order 21~cm bias factors, $b_{21}$ and $b_{21}^{(2)}$. (For brevity of notation, we
do not include a super-script $^{(1)}$ in our notation for the first-order bias factors.) 
The Fourier space counterpart of Eq.~\ref{eq:bias_expand} is,
\beq
\label{eq:bias_fourier}
\delta_{21}(\k) = b_{21} \delta_\rho(\k) + \frac{b_{21}^{(2)}}{2} \int \frac{d^3q}{(2 \pi)^3} \delta_\rho(\q) \delta_\rho(\k -\q),
\eeq 
i.e. the $\delta^2_{\rho}(\x)$ term in configuration space becomes a convolution in Fourier space. 
In practice, since
$\avg{T_{21}}$ is not directly observable in an interferometric measurement, we generally consider the bias factors multiplied by $\avg{T_{21}}$.
Under the assumption of a local and deterministic biasing relation for the [CII] specific intensity field, we define analogous bias factors
$b_{\CII}$ and $b_{\CII}^{(2)}$.

As large-scale structure grows under gravity, the matter density field will develop non-Gaussianity. At second-order in perturbation theory, the density fluctuations
$\delta_\rho(\k)$ follow (e.g. \citealt{Bernardeau:2001qr}):
\beq \label{eq:secondorder_delta}
\delta_\rho(\bm{k}) = \delta_\rho^{(1)} (\k) + \int \frac{\text{d}^3q}{(2\pi)^3} F_2(\bm{q},\bm{k}-\bm{q})\delta_\rho^{(1)}(\bm{q})\delta_\rho^{(1)}(\bm{k}-\bm{q}),
\eeq
where $\delta_\rho^{(1)}$ denotes the first-order fluctuation field and $F_2(\k_1,\k_2)$ describes mode-coupling from non-linear growth. In a flat $\Lambda$CDM universe with $\Omega_m(z) \approx 1$, appropriate for high redshifts, the mode-coupling kernel is given by:
\beq \label{eq:F2_eq}
F_2 (\bm{q}_1,\bm{q}_2) = \frac{5}{7} + \frac{\bm{q_1} \cdot \bm{q}_2}{2q_1q_2}\left(\frac{q_1}{q_2} + \frac{q_2}{q_1} \right) + \frac{2}{7} \left( \frac{\bm{q}_1 \cdot \bm{q}_2}{q_1q_2}\right)^2.
\eeq
This mode-coupling sources the matter density bispectrum, which is defined by:
\beq \label{eq:bispect}
\begin{split}
\langle \delta_\rho(\bm{k}_1)\delta_\rho(\bm{k}_2)\delta_\rho(\bm{k}_3) \rangle &\equiv  (2\pi)^3 \delta_D(\bm{k}_1 + \bm{k}_2 + \bm{k}_3) \\
& \times B_{\delta, \delta, \delta}(\bm{k}_1,\bm{k}_2,\bm{k}_3)\text{.}
\end{split}
\eeq
Using the result of second-order perturbation theory, Eqs.~\ref{eq:secondorder_delta} and \ref{eq:F2_eq} gives the bispectrum to lowest non-vanishing order as:
\beq \label{eq:bispec_den}
\begin{split}
B^{(0)}_{\delta,\delta,\delta}(\bm{k}_1,\bm{k}_2,\bm{k}_3) = & 2 F_2 (\bm{k}_1,\bm{k}_2)P_{\delta, \delta}^{\text{lin}}(k_1)P_{\delta, \delta}^{\text{lin}}(k_2) \\
& + \text{2 perm.},
\end{split}
\eeq
where $P^{\text{lin}}_{\delta, \delta}$ is the linear matter auto-spectrum and $\text{2 perm.}$ indicates permutations of the products of power spectra that
enter, with each permutation picking out a product of two of the three wavevectors involved. We make a distinction here between the linear and nonlinear matter power spectra, but there is only a $\sim5\%$ difference between the two for the smallest scales ($k \sim 0.4\,\Mpc^{-1}$) and lowest redshift ($z \sim 6$) we consider.
Since $B \propto P^2$, it is also convenient to define the
reduced bispectrum. In the case of the matter density field,
\beq \label{eq:qred_den}
Q_{\delta, \delta, \delta}(\bm{k}_1,\bm{k}_2,\bm{k}_3) =  \frac{B_{\delta, \delta, \delta}(\bm{k}_1,\bm{k}_2,\bm{k}_3)}{P_{\delta, \delta}^{\text{nl}}(k_1)P_{\delta, \delta}^{\text{nl}}(k_2) + \text{2 perm.}}.
\eeq
It is common to specify $Q$ by the magnitude of wavevectors, $k_1 = |\bm{k}_1|$, and $k_2=|\bm{k}_2|$, and the angle between $\bm{k}_1$ and $\bm{k}_2$, $\theta_{12}$.\footnote{We will use a different Fourier space characterization when we actually extract $b_{21}$ from our simulations, see \S~\ref{ssec:bispectrum_algorithm}}
The reduced matter bispectrum has a distinctive form: first, gravitational mode-coupling (Eqs.~\ref{eq:secondorder_delta}-\ref{eq:qred_den}) enhances 
the small-scale power spectrum in large-scale overdense regions compared to that in underdense regions.
Second, the reduced bispectrum is smaller for triangles that are close to isosceles with $\theta_{12} \sim \pi/2$, than for triangles that are nearly co-linear with $\theta_{12} \sim 0, \pi$. This
is a consequence of the filamentary nature of large-scale structure, and leads to a characteristic ``U''-shaped dependence of $Q_{\delta, \delta, \delta}$ on $\theta_{12}$ (e.g. \citealt{Bernardeau:2001qr}). 

Turning to biased-tracers, such as the 21~cm or [CII] fluctuations, it is clear that (at lowest non-vanishing order) the bispectra of these fields receive contributions
both from non-Gaussianity in the matter distribution (described by Eq.~\ref{eq:qred_den}) and owing to non-linearities in the biasing relation (Eqs.~\ref{eq:bias_expand}-\ref{eq:bias_fourier}).
Here we consider the cross-bispectrum between the 21~cm field and two copies of the [CII] fluctuation field. This quantity is less sensitive to foreground contamination -- and to unshared systematics -- than the auto-bispectra of these fields. 
This is because the foregrounds in the two surveys should be mostly uncorrelated (asides for common foregrounds at the widely separated observing frequencies of the two experiments), and
so should not contribute to the ensemble-averaged cross-bispectrum.\footnote{Any correlations between the signal in one survey and the foregrounds in the other could also produce a small ensemble-averaged cross-bispectrum. For example, the high redshift [CII] emitting galaxies -- or other correlated sources at the redshift of these galaxies -- may emit synchrotron radiation which constitutes
a (very small) ``foreground'' for the 21~cm survey.}
 Foreground cleaning or avoidance is still important because residual foregrounds will increase the variance of a cross-bispectrum estimate, but the cross-bispectrum has the important virtue that unshared foregrounds will not produce a spurious signal {\em on average}.  The cross-power spectrum is also less
sensitive to foreground contamination; however on large scales, the cross-spectrum is proportional to the overall product $b_{21} \avg{T_{21}} b_{\CII} \avg{I_{\CII}}$. That is, the cross-power spectrum depends on both the 21~cm and [CII] biasing relations, as well as the mean intensity in each line. On the other hand, we will show that the shape of the cross-bispectrum -- when suitably defined -- is sensitive 
only to $\bto$ (and independent of the [CII] biasing and mean intensity). 

Specifically, consider the cross-bispectrum between a single 21~cm field and two [CII] emission fields:
\beq \label{eq:bispec_cross}
\begin{split}
\langle T_{21} (\bm{k}_1) I_{\CII}(\bm{k}_2) I_{\CII}(\bm{k}_3) \rangle \equiv & \,(2\pi)^3 \delta_D(\bm{k}_1 + \bm{k}_2 + \bm{k}_3) \\
& \times B_{21, \CII, \CII}(\bm{k}_1,\bm{k}_2, \bm{k}_3).  
\end{split}
\eeq
Note that we have now switched from considering $\delta_{21}$ and $\delta_{\CII}$ to $T_{21}$ and $I_{\CII}$, since only the latter quantities are directly observable from interferometric and line-intensity mapping measurements.
Apart from the overall $\bm{k}=0$ mode, $T_{21}(\bm{k}) = \avg{T_{21}} \delta_{21}(\bm{k})$, and $I_{\CII}(\bm{k}) = \avg{I_{\CII}} \delta_{\CII}(\bm{k})$. 

We can define a reduced cross-bispectrum that is formed from Eq.~\ref{eq:bispec_cross} and the cross-power spectrum between the 21~cm and [CII] fields, $P_{21,\CII}(k_i)$, at each of the three
wavevectors $k_i$ with $i=1,2,3$.  
Specifically,
\beq \label{eq:red_crossbisp}
\hat{Q}_{21, \CII, \CII}(\bm{k}_1,\bm{k}_2,\bm{k}_3) = \frac{B_{21, \CII, \CII}(\bm{k}_1,\bm{k}_2,\bm{k}_3)}{P_{21, \CII}(k_1)P_{21, \CII}(k_2) + \text{2 perm.}}
\eeq
Note that we deliberately construct the reduced cross-bispectrum using only the cross-power spectrum in the denominator of Eq.~\ref{eq:red_crossbisp}: explicitly, ``$\text{2 perm.}$'' is specified
by $P_{21,\CII}(k_2) P_{21,\CII}(k_3) + P_{21,\CII}(k_3) P_{21,\CII}(k_1)$ and does not involve the auto-power spectrum of either field. 
Similar definitions hold for $\hat{Q}_{\CII, 21, \CII}$ and $\hat{Q}_{\CII, \CII, 21}$; these merely rearrange which wavevector is attached to the 21~cm field. Note that with this definition, using ${\rm mK}$ units
for $T_{21}$ and ${\rm Jy/str}$ units for $I_{\CII}$, the cross-bispectrum $B_{21, \CII, \CII}$ has units of ${\rm mK} ({\rm Jy/str})^2 ({\rm Mpc})^6$, while the reduced cross-bispectrum $\hat{Q}_{21, \CII, \CII}$ has units of ${\rm mK}^{-1}$. 
\begin{gather*} \label{eq:rcb_lowest}
\hat{Q}_{21, \CII, \CII}(\bm{k}_1,\bm{k}_2,\bm{k}_3) = \frac{Q_{\delta, \delta, \delta}(\bm{k}_1,\bm{k}_2,\bm{k}_3)}{\bto} \\
 + \frac{b_{21}^{(2)}}{2\avg{T_{21}}b_{21}^2} L(k_1,k_2) \\
 + \frac{b_{\CII}^{(2)}}{2\avg{T_{21}}b_{21}b_{\CII}}\big[ L(k_2,k_3) + L(k_3,k_1)\big]\text{,}
\end{gather*}
where we have defined
\beq \label{eq:loser_term}
L(k_i, k_j) = \frac{P_{1,2}(k_i)P_{1,2}(k_j)}{P_{1,2}(k_1)P_{1,2}(k_2) + \text{2 perm.}}\text{.}
\eeq
As will become clear shortly, these $L$ terms are inconvenient, but note that
\beq \label{eq:bye_losers}
L(k_1,k_2) + L(k_2,k_3) + L(k_3, k_1) = 1\text{.}
\eeq
Therefore, summing permutations of Eq.~\ref{eq:rcb_lowest} and using Eq.~\ref{eq:bye_losers} rids us of the $L$ terms:
\beq \label{eq:crossbisp_def}
Q_{21, \CII, \CII} \equiv \frac{1}{3} \left( \hat{Q}_{21, \CII, \CII} + \hat{Q}_{\CII, 21, \CII} + \hat{Q}_{\CII, \CII, 21} \right)\text{.}
\eeq
Throughout we will only work with $Q_{1, 2, 2}$, and so these permutations are implicit in what follows. Note that the permutations in Eq.~\ref{eq:crossbisp_def} imply that swapping any two of the wavevector arguments results in the same value of $Q$. Therefore, we will also implicitly enforce $k_1 \geq k_2 \geq k_3$ where relevant, to avoid double counting.

With these permutations in mind, we now have the following formula:
\beq \label{eq:crossbisp}
\begin{split}
Q_{21, \CII, \CII}=& \frac{Q_{\delta, \delta, \delta}}{\bto} + \frac{1}{6}\frac{b_{21}^{(2)}}{\avg{T_{21}}b_{21}^2} \\
&+ \frac{1}{3}\frac{b_{\CII}^{(2)}}{\avg{T_{21}}b_{21}b_{\CII}} \text{.}
\end{split}
\eeq
As we have described, this formula is valid at second-order in perturbation theory.
In our actual analysis, we will use the simulated $Q_{\delta,\delta,\delta}$ instead of the perturbative $Q_{\delta,\delta,\delta}$. This is mostly to minimize the impact of sample variance and is discussed further in \S~\ref{sec:results}.

Our analysis strategy is now clear, and analogous to methods developed previously to constrain galaxy bias \citep{Fry94,Matarrese:1997sk,Scoccimarro:2000sp,Verde:2001sf}. The first term in $Q_{21, \CII, \CII}$ depends on the reduced density bispectrum, with its distinctive ``U''-shaped dependence on $\theta_{12}$, and the 21~cm linear bias factor. In this term, only the 21~cm bias enters and so this term is entirely independent of the statistical properties of the [CII] emission fluctuations. 
In contrast, the other terms depend on the first and second-order bias factors of both the 21~cm and [CII] fields, and so these factors are harder to interpret, but produce only a constant offset. The rationale for taking the permuted version (Eq.~\ref{eq:crossbisp_def}) is now clarified, since as a result the [CII] bias factors enter only as an overall constant, independent of triangular configuration. In summary, by measuring the triangular shape dependence of $Q_{21, \CII, \CII}$, we should be able to extract $\bto$.

In testing the perturbative framework described above, we will consider various additional bispectra. At lowest non-vanishing order in perturbation theory, these all have the general form:
\beq \label{eq:Q21XX}
Q_{21,\text{X},\text{X}}^{(0)} = \frac{Q_{\delta, \delta, \delta}}{\bto} + C_{21,\text{X},\text{X}},
\eeq
where the various values of $C$ depend on which field ``X'' is being combined with the 21~cm fluctuations. Specifically,
\beq \label{eq:C212121}
C_{21,21,21} = \frac{1}{2}\frac{b_{21}^{(2)}}{\avg{T_{21}}b_{21}^2},
\eeq
\beq \label{eq:C21deldel}
C_{21,\delta,\delta} = \frac{1}{6}\frac{b_{21}^{(2)}}{\avg{T_{21}}b_{21}^2},
\eeq
and
\beq \label{eq:C21CIICII}
C_{21,\text{CII},\text{CII}} = \frac{1}{6}\frac{b_{21}^{(2)}}{\avg{T_{21}}b_{21}^2} + \frac{1}{3}\frac{b_{\text{CII}}^{(2)}}{\avg{T_{21}}b_{21}b_{\text{CII}}}\text{.}
\eeq

\section{Reionization Simulations and Methodology} \label{sec:methods}
In order to characterize the cross-bispectrum signal at different stages of reionization, and to test the accuracy of the perturbative formulas from the previous section, we turn to semi-numerical simulations of the EoR \citep{Zahn:2006sg,Mesinger:2007pd}, based on the excursion-set \citep{Bond:1990iw} model of reionization \citep{Furlanetto:2004nh}. Specifically, we use the publicly available 
\texttt{21cmFAST} code v1.12 \citep{Mesinger11}.

Our simulations are unlikely to provide fully accurate models of the small-scale 21~cm signal. In the context of this paper, the small-scale modes mostly set the overall values of the bias factors ($\bto$) at different redshifts and neutral fractions. The small-scale modes should therefore impact the precise evolution of the bias factors with redshift, and not the overall framework we propose here.

\subsection{L-PICOLA}\label{ssec:lpicola}
By default, \texttt{21cmFAST} uses the Zel'dovich approximation (ZA) to generate the density field \citep{ZA1970}.
This is appropriate for capturing the large-scale two-point statistics of the 21~cm field at high redshifts, but it is inadequate for modeling three-point statistics. The ZA is known to underestimate the density bispectrum \citep{1997cosccimarro:cosm_pert,2013leclercq}, and we have verified that it provides a poor description of the density bispectrum at the scales and redshifts of interest for our study. 
We thus turn to the publicly available \texttt{L-PICOLA} v1.3 code \citep{Howlett:2015lpicola,Tassev2013:COLA}, which is a hybrid between a 2nd order Lagrangian perturbation theory (2LPT) and a particle mesh (PM) code. \texttt{L-PICOLA} is a parallel version of the \texttt{COLA} method, and accurately reproduces the two and three-point statistics of interest of our study, yet is less computationally expensive than full GADGET-2 \citep{Springel:2005mi} runs.

In order to accurately estimate the bispectra starting at $z \sim 10.5$, two potential issues are: transients associated with the initial conditions \citep{Crocce:2006transients} and discreteness noise from
the finite number of particles in the simulation. These issues are more of a concern at the high redshifts of interest for our study (than near $z=0$), since there is less time for transients to relax and because the density fluctuations are small at early times. The transients are suppressed in part by adopting 2LPT initial conditions (as opposed to Zel'dovich initial conditions) and by starting the simulation early. We guard against shot-noise by simulating a fairly high particle density. Specifically, we simulate $1024^3$ particles in a simulation box with a co-moving side length of
$L_{\rm box} = 800\, \Mpc$. We chose an extremely high starting redshift of $z_i=5000$.\footnote{The starting redshift is artificially high; the initial conditions are set by rescaling from the $z=0$ matter power spectrum assuming that the universe is dominated by pressure-free matter at the initial redshift (even though this redshift is formally before matter-radiation equality.) The subsequent
evolution is tracked by \texttt{L-PICOLA} and this returns the correct linear evolution (under the approximation that the universe is entirely composed of pressure-free matter) after recombination. The early start was just meant to guard against initial transients and to ensure the correct density statistics at $z \sim 6-11$.} The code takes 300 times steps between $z_i$ and $z=10.49$.

In order to beat-down noise in our bispectrum estimates on large scales, we run ten different realizations of the \texttt{L-PICOLA} simulations, resulting in a total simulated volume of $5.12 \, \Gpc^3$.
The boxsize was chosen so that the fundamental mode of the simulation box is much larger than the scales of interest for our study. Specifically, the largest scale considered is $k=0.05\, \Mpc^{-1}$ and the fundamental mode of the simulation box is  $k_f = 0.00785\,\Mpc^{-1}$. We save simulation snapshots at redshifts of $z=10.49$, $9.41$, $8.43$, $7.88$, $7.37$, $7.05$, $6.73$, $6.43$, and $6.00$, and take ten time steps between each output. For each redshift snapshot, we use nearest grid-point interpolation to estimate the density field on a $512^3$ Cartestian grid from the particle positions.

\subsection{21cmFAST}\label{ssec:21cmFAST}

The $512^3$ particle, $L_{\rm box} = 800\, \Mpc$, gridded density fields from the \texttt{L-PICOLA} runs are then passed to the \texttt{21cmFAST} code to generate simulated ionization and 21~cm fields
at each redshift. Specifically, the excursion set approach is used to identify ionized regions \citep{Mesinger11} using  the following simulation parameters: the ionizing efficiency is set to $\zeta=10$,
the minimum virial temperature of galaxy hosting halos is $T_{\text{vir}} = 10^4$ K; the largest smoothing scale -- for generating the ionization field using the excursion-set methodology -- is $R_{\text{max}}=30$ co-moving Mpc. (The latter quantity is sometimes loosely termed the ``mean-free path'', $R_{\text{mfp}}$, in the literature.)

\begin{figure}
\includegraphics{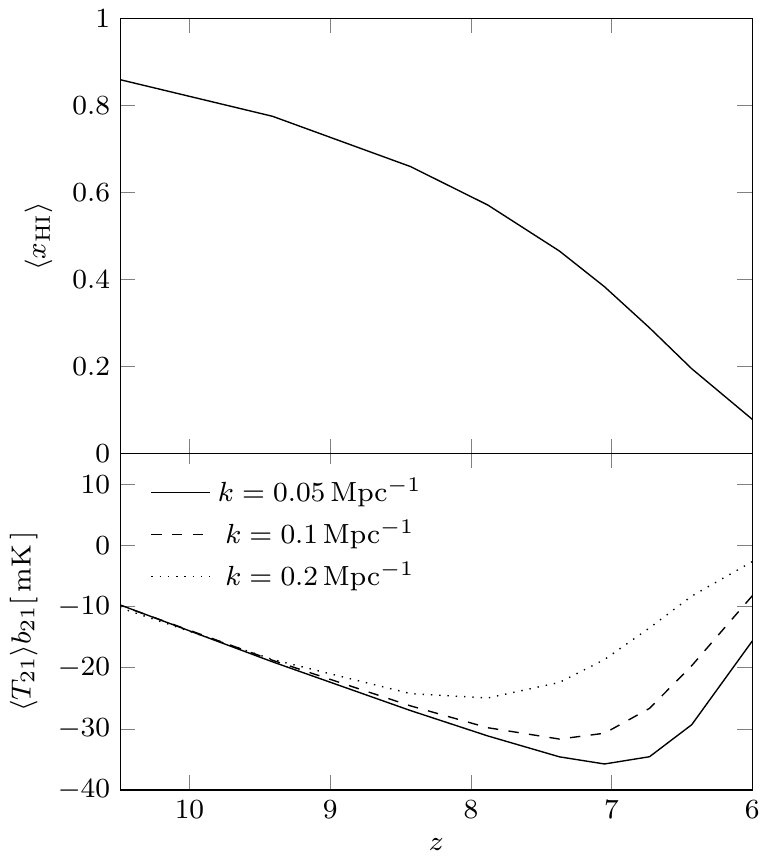}
\caption{The ionization history and the redshift evolution of the 21~cm bias factor. {\em Top panel}: The volume-averaged neutral hydrogen fraction, $\avg{x_{\HI}}$, as a function of redshift from our 21cmFAST
simulation. {\em Bottom panel}:  The corresponding evolution of the 21~cm bias factor, $\bto$, which rises and falls with decreasing redshift and peaks (in amplitude) around reionization's midpoint, when $\avg{x_i} \sim 0.5$ -- or slightly thereafter -- 
of the IGM volume is reionized. We compute $\bto$ from the 21~cm-density cross-spectrum at various $k$. At high redshifts (early in reionization) $\bto$ is insensitive to the precise wavenumber
considered, but differences appear as reionization proceeds. This provides one indication of the redshifts and scales where the linear biasing description becomes imperfect. 
}
\label{fig:ionization_history}
\end{figure}

The {\em Top panel} of Fig.~\ref{fig:ionization_history} shows the redshift evolution of the volume-averaged neutral fraction, $\avg{x_{\HI}}$, in our 21cmFAST model.  In this model, reionization completes shortly after $z \sim 6$,  while $50 \%$ of the volume is ionized at $z \sim 7.5$, and $20 \%$ is ionized by $z \sim 10$; this is broadly consistent with current constraints on the EoR 
(e.g. \citet{2001becker:reion_constraint,2006fan:reion_constraint,2013venemans:reion_constraint,2015becker:reion_constraint,2016planck:reion_constraint}. )
The {\em Bottom panel} shows the corresponding evolution of the 21~cm bias factor with redshift, $\bto$ (estimated from the 21 cm density cross-power spectrum as we detail subsequently) calculated at each
of $k=0.05\, \Mpc^{-1}, 0.1\, \Mpc^{-1}$, and $0.2\, \Mpc^{-1}$.  Fig.~\ref{fig:ionization_history} exemplifies the distinctive ``rise and fall'' evolution in the magnitude of $\bto$: a key goal of upcoming 21~cm surveys is to measure this evolution, and to use it to extract the corresponding ionization history. Reiterating, the motivation of our paper is to develop an approach for extracting this evolution which is less susceptible to foreground contamination than the 21~cm auto spectrum.
At early times, the bias factor inferred is identical for all three spatial scales considered. However, as reionization proceeds the bias factor starts to depend on wavenumber: this occurs because
the ionized regions grow in size during reionization, and the linear biasing description breaks down on progressively larger scales. Hence, even on the rather large scales spanned by our simulation box we expect linear biasing to be an imperfect description during the late stages of reionization (as we will quantify further subsequently). In practice, it will probably be challenging to access still larger scales from future surveys (where linear biasing might be a better description), since foreground avoidance or cleaning limit the prospects for extracting very large scale modes. (See, e.g., Fig. 1 of \citealt{Dillon:2013rfa} and the discussion therein.)

Lastly, note that the bias factor in the {\em Bottom panel} of Fig.~\ref{fig:ionization_history} is {\em negative}: this indicates that large-scale overdense regions are dimmer in 21~cm emission than
typical regions. The negative sign results because the ionizing sources form first in large-scale overdense regions and so these regions are ionized early. It is worth remarking that the 21~cm auto-power spectrum is sensitive only to $\left[\bto\right]^2$ and hence can not distinguish the overall sign here. On the other hand, the 21~cm-[CII] cross-spectrum, the 21~cm auto-bispectrum, and the 21~cm-[CII]-[CII] cross-bispectrum each depend on both the magnitude and the sign of $\bto$, as has also been shown by \citet{Majumdar:2017tdm,Hoffman2018:2point3point}.

\begin{figure}
\includegraphics{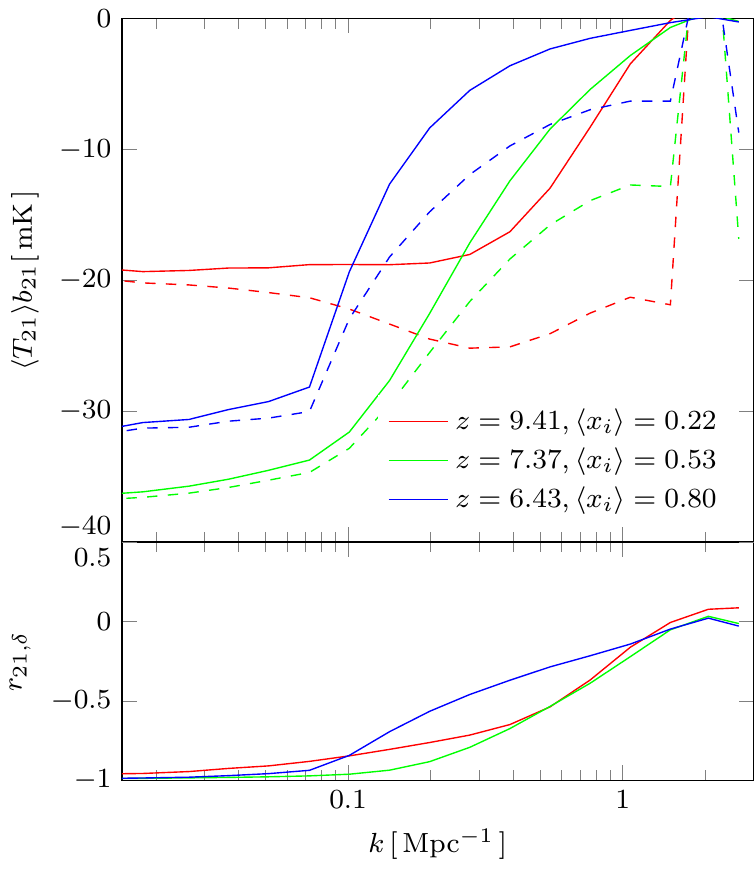}
\caption{Linear biasing of the 21~cm fluctuations and its range of validity.
{\em Top panel}: Estimates of the scale-dependent bias factor, $\bto(k)$, from our reionization simulation for several different redshifts and ionization fractions ($\avg{x_i}$), as described in the legend. 
The solid lines show estimates from the 21~cm auto-power spectrum, while the dashed lines show the bias factor inferred from the 21~cm-density cross-power spectrum.
{\em Bottom panel}: The cross-correlation coefficient between the 21~cm and matter density fields, $r_{21, \delta} (k)$. At $k \lesssim 0.1\,\Mpc^{-1}$, the bias factor is close to scale-independent, 
and $r_{21,\delta}(k) \sim -1$.
}
\label{fig:b21_fromps}
\end{figure}

We can carry out further tests of the linear biasing approximation using additional two-point statistical measures. As already mentioned, we expect this approximation to break down on sufficiently small scales, especially during the later stages of reionization. 
On scales smaller than the size of the ionized regions, there are order unity fluctuations in the ionization fraction, as some areas in the IGM are nearly completely ionized while others remain almost completely neutral. In addition, the ionization and density field de-correlate on scales smaller than the size of the ionized bubbles (e.g. \citealt{Furlanetto:2004nh,Zahn:2006sg}). Hence on these scales, the 21~cm and density fluctuation fields do not share the same phases and an expansion along the lines of Eq.~\ref{eq:bias_expand} is inadequate. To explore this we estimated the bias factor from the cross-power spectrum of the 21~cm and density fields, $b_{21,x}(k) = P_{21, \delta}(k)/P_{\delta, \delta}(k)$, and from the 21~cm auto-power spectrum $b_{21, a}(k) = \left[P_{21, 21}(k)/P_{\delta, \delta}(k)\right]^{1/2}$ ({\em Top panel} of Fig.~\ref{fig:b21_fromps}).  Note that these two estimates should agree in the linear-biasing regime, but may depart from each other on smaller scales (at higher $k$). Furthermore, we calculate the cross-correlation coefficient between the 21~cm and density fields, $r_{21,\delta}(k)=P_{21,\delta}(k)/\sqrt{P_{21,21}(k) P_{\delta,\delta}(k)}$ ({\em Bottom panel}). The cross-correlation coefficient quantifies how well the phases of the two fields track each other.

At $k \lesssim 0.1\,\Mpc^{-1}$, for the redshifts shown: the correlation-coefficient between the two fields is $r_{21,\delta} \sim -1$, and the bias factor is fairly scale-independent. The auto/cross-power spectrum approaches for calculating the bias factor also agree with each other to within $\sim 20\%$ for $k \sim 0.1 \,\Mpc^{-1}$, and to within $\sim 5\%$ for $k \sim 0.05\,\Mpc^{-1}$ at $\avg{x_i}=0.8$ ($z=6.43$). The linear biasing description appears best in the middle of reionization. Towards the end of reionization, as illustrated by the blue $\avg{x_i}=0.80$ curves, the bubbles are large enough that $k \sim 0.1\,\Mpc^{-1}$ is only marginally in the linear biasing regime. On the other hand, in the earliest phases of reionization ($\avg{x_i} \sim 0.1$), the
correlation-coefficient departs significantly from $r_{21,\delta} \sim -1$: this likely results because the sign of $r_{21,\delta}$ is initially {\em positive} and gradually reverses \citep{Lidz08}. 
This occurs because the large-scale overdense regions prior to reionization contain more neutral hydrogen and are brighter in 21~cm emission than underdense areas, while the large-scale overdense regions ionize first and subsequently become dimmer in 21~cm. In our model large-scale overdense and underdense regions are nearly at the same brightness temperature at 
$\avg{x_i} \sim 0.1$ while the reversal is almost complete by $\avg{x_i} = 0.22$ at $z=9.41$ (the earliest case shown in Fig.~\ref{fig:b21_fromps}). We do not expect our perturbative formula to be accurate in the brief earlier phase near $\avg{x_i} \sim 0.1$, where $r_{21,\delta}$ changes sign rapidly. Note also that we neglect spin temperature fluctuations, which should impact our predictions at early times.
In general, however, our results suggest that linear biasing is a fairly good description at $k \lesssim 0.1\,\Mpc^{-1}$ during the bulk of reionization, further motivating us to test our cross-bispectrum method.

\subsection{[CII] Intensity Field} \label{ssec:CII_model}

In order to model the [CII] intensity field, our simulations would ideally resolve halos with masses down to the atomic cooling limit ($\sim 10^8\,M_{\sun}$). In this case, we could model the high redshift [CII] intensity field by assuming correlations between halo properties and [CII] luminosity. However, it is difficult to resolve such small mass halos while capturing a large enough volume to accurately model the bispectrum. For example, the particle mass in our simulations is $\sim 10^{10} M_{\sun}$ and so we don't resolve many of the small mass halos which may host [CII] emitting galaxies. Consequently, we defer a full treatment of [CII] emission to future work.

Note, however, that the perturbative formulae (Eq.~\ref{eq:crossbisp} and Eq.~\ref{eq:Q21XX}) predict -- apart from the overall constant term -- that the $Q_{21,\CII,\CII}$ bispectrum should be identical to that of the $Q_{21,\delta,\delta}$ bispectrum. That is, both of these cross-bispectra are expected to follow the $Q_{\delta,\delta,\delta}/(\bto)$ form. Consequently, we study the statistics of $Q_{21,\delta,\delta}$ here rather than $Q_{21,\CII,\CII}$ since the latter is more difficult to model. The main shortcoming of this approach is that it neglects the additional non-linearities in the [CII] biasing relation, and these effects may reduce the  agreement with the perturbative formula.
In estimating the signal-to-noise ratio at which $Q_{21,\CII,\CII}$ may be measured in future surveys (\S~\ref{sec:detectability}), we assume a linear biasing model for [CII] emission. This allows us to predict the [CII] auto-spectrum and [CII]-21 cm cross-spectrum (as required for the signal-to-noise ratio calculation) from the statistics of the simulated density field.
It may be possible to construct a sub-grid model to include unresolved small halos with the correct statistical properties (e.g. \citet{2007mcquinn:lowmasshalos}). However, the approach adopted in that work would need to be extended to capture the correct quadratic halo bias and is beyond the scope of this paper.

\subsection{Estimating the Bispectrum} \label{ssec:bispectrum_algorithm}

In order to estimate the bispectrum from our 21cmFAST simulation outputs, we follow the method described in \citet{Smith:2007sb}. Throughout we consider two characterizations of the wavevector arguments. In both cases we calculate the average bispectrum in linearly-spaced wavevector bins.

Our first characterization is for illustration purposes only. Here we fix the magnitude of $|\k_1|$ and $|\k_2|$ while varying the angle $\theta_{12}$ between the two wavevectors. The third side of the triangle, $\k_3$, is determined by these three parameters and the requirement of translation invariance ($\k_1 + \k_2 + \k_3 = \bm{0}$.) We consider spherical bins around each wavevector of thickness $\Delta k = k_f$, where $k_f = 2\pi/L_{\text{box}} = 0.00785\,\Mpc^{-1}$ is the fundamental mode of our simulation box.

Our second characterization sets a minimum ($k_{\tmin}$) and a maximum ($k_{\tmax}$) magnitude for each wavevector argument. That is, we enforce $k_{\tmin} < |\k_i| < k_{\tmax}$ for each wavevector $\k_i$, along with the usual constraint that $\k_1 + \k_2 + \k_3 = \bm{0}$. Owing to homogeneity and isotropy, all bispectra have three arguments: $|\k_1|$, $|\k_2|$, $|\k_3|$. Our bins are cubes of side length $4\,k_f$ in Fourier space. This characterization captures nearly all the available modes. As discussed in \S~\ref{sec:approach}, we also enforce that $k_1 \geq k_2 \geq k_3$.
In estimating bispectra, we loop through all of the simulated Fourier modes that lie within this wavevector range. We determine $B_{21, \delta, \delta}$ and $P_{21, \delta}$ from the same set of Fourier modes to reduce sample variance in our estimates of $Q_{21,\delta,\delta}$. In general, one would like to make $k_{\tmin}$ and $k_{\tmax}$ as small and large as possible, respectively. However,
our estimates at small $k$ are noisy and the perturbative formula becomes less accurate at high $k$.
We chose values of $k_{\tmin}$ and $k_{\tmax}$ throughout that give the best results, but also show results for different values of $k_{\tmin}$ and $k_{\tmax}$ in Appendix A.

\section{Results}\label{sec:results}

\begin{figure}
\includegraphics{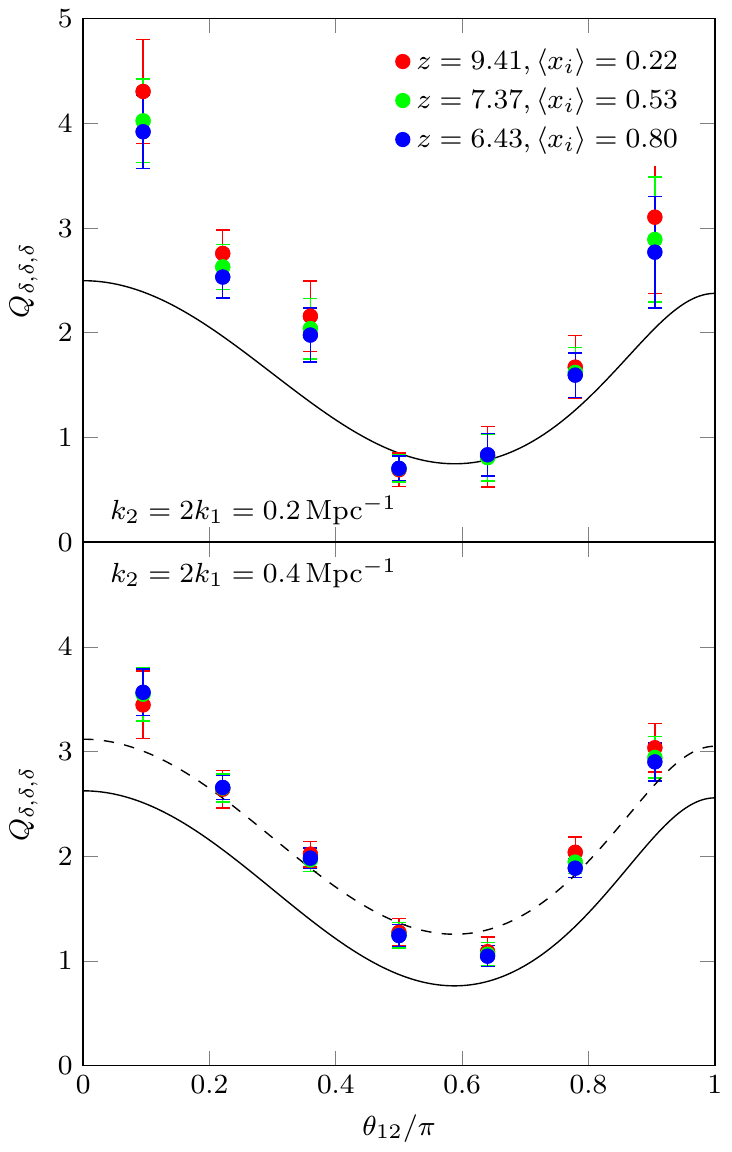}
\caption{Reduced bispectrum of the density field. Here we fix $k_2 = 2k_1 = 0.2\,\Mpc^{-1}$ in the {\em Upper panel} and $k_2 = 2k_1 = 0.4\,\Mpc^{-1}$ in the {\em Bottom panel}. We plot the results as a function of $\theta_{12}$, the angle between the two wavevectors. The solid line is the prediction of second-order perturbation theory, as given in Eq.~\ref{eq:F2_eq}. Each colored point shows the average reduced bispectrum from ten different simulation realizations. The error bars give estimates of the error on the average, as determined from the scatter across the ten simulation realizations. Note that the poorer sampling for nearly collinear triangle configurations leads to noisier estimates near 
$\theta_{12}/\pi \approx 0\text{,} 1$. The dashed line is the perturbative $Q_{\delta,\delta,\delta}$ but with a constant added to minimize the $\chi^2$ difference with the $z=7.37$ measurements. This indicates that, for the most part, the perturbative $Q_{\delta,\delta,\delta}$ differs from the simulation results only by an overall constant.}
\label{fig:Qdeldeldel}
\end{figure}

First we consider the reduced auto-bispectrum of the simulated matter density field, $Q_{\delta,\delta,\delta}(\theta_{12})$, at the redshifts of interest in Fig.~\ref{fig:Qdeldeldel}. As discussed in \S~\ref{ssec:lpicola}, we average over ten independent simulation realizations. 
We show the simulated reduced density bispectrum for triangles with $k_1=0.1\,\Mpc^{-1}$ and $k_2=2 k_1$ and a case where the wavenumbers are twice as large. In the larger-scale case, the simulated bispectra are somewhat noisy while the smaller-scale case provides a smoother bispectrum estimate, because of the larger number of sampled modes at smaller scales. The simulation measurements are generally similar to the predictions of second order perturbation theory (Eq.~\ref{eq:bispec_den}) for the large scales shown and the high redshifts considered, although there are noticable differences.
The departures from the predictions of second-order perturbation theory evident in the figure reflect non-linearities that are not fully captured at second-order in $\delta$. In order to partly account for these non-linear density fluctuations, we subsequently use the {\em simulated}
density auto-bispectrum, $Q_{\delta,\delta,\delta}$, as input to the perturbative formulae (Eqs.~\ref{eq:crossbisp} and \ref{eq:Q21XX}). Throughout, we refer to this prediction loosely as the ``second-order perturbation theory prediction'' although strictly speaking it accounts for the impact of additional non-linearities in the form of $Q_{\delta,\delta,\delta}$. This also serves to reduce the impact
of sample variance in our comparisons between simulations and perturbation theory estimates.

That being said, the departures from non-linearity manifest mainly as an overall constant. We illustrate this by showing in Fig.~\ref{fig:Qdeldeldel} (dashed line) the perturbative $Q_{\delta,\delta,\delta}$ but with a constant added to minimize the $\chi^2$ statistic at $z=7.37$. The agreement between the dashed line and the points at $k_2 = 2k_1 = 0.4\,\Mpc^{-1}$ indicates that the non-linearities will propagate into $C_{21,\text{X},\text{X}}$ but not into $\bto$. While this is not the case for the noisier $k_2 = 2k_1 = 0.2\,\Mpc^{-1}$ panel, we nonetheless proceed with these caveats in mind.

\begin{figure*}
\centering
\includegraphics{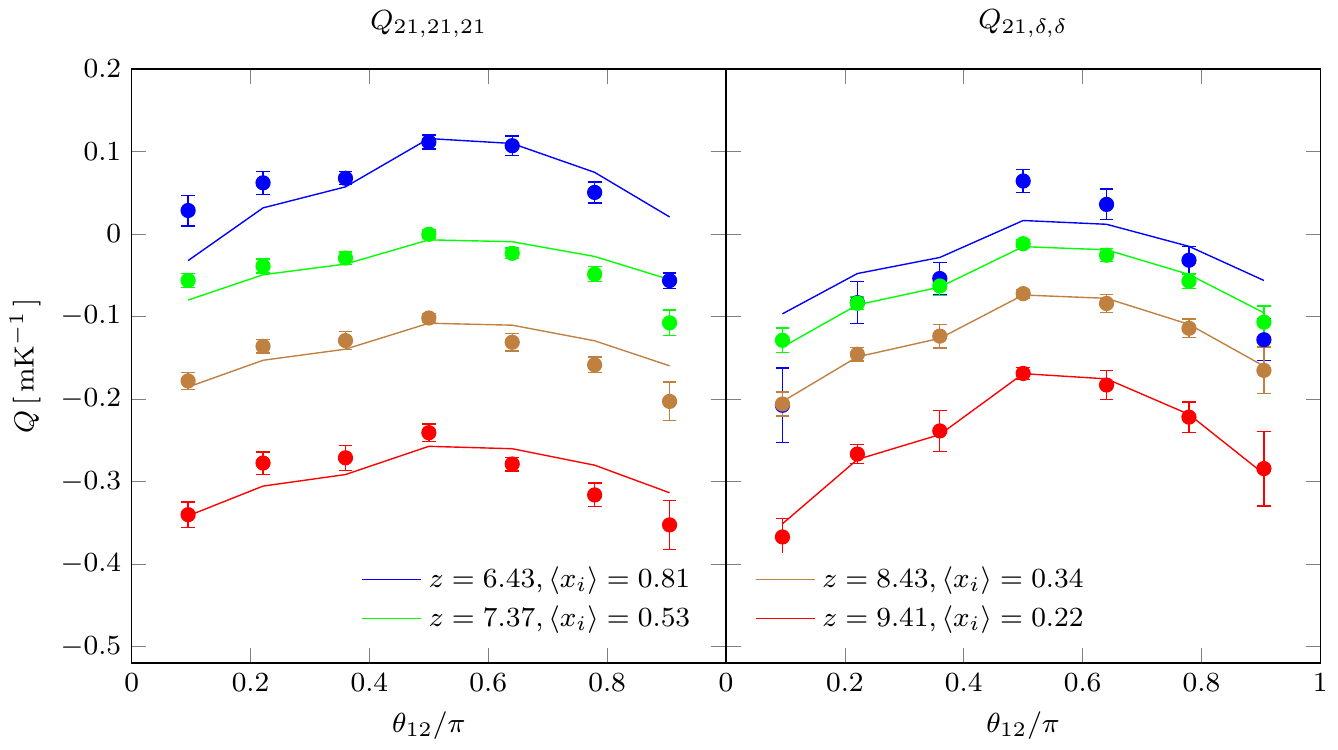}
\caption{Evolution of various reduced bispectra with redshift/ionization fraction. {\em Left}: the reduced 21~cm auto-bispectrum. {\em Right}: the reduced 21~cm-density-density cross-bispectrum. In each panel, the solid lines show the best fit models of the form $Q_{\delta,\delta,\delta}/\bto + C$, using the simulated values of $Q_{\delta,\delta,\delta}$. We restrict ourself to modes that satisfy $k_1 = 0.1\,\Mpc^{-1}$ and $k_2 = 0.2\,\Mpc^{-1}$, varying the angle $\theta_{12}$ between the two wavevectors. Note that the best fit values of $\bto$ and $C$ here are obtained by
fitting only to this restricted range of triangle configurations and differ somewhat from the best fits obtained subsequently, which match to a broader range of triangles. The perturbative formula provides a better match to $Q_{21,\delta,\delta}$ than $Q_{21,21,21}$ as we discuss in the text. Note that since we use the simulated $Q_{\delta,\delta,\delta}$ in our fits, the curves share some of the noise in our simulation estimates.} 

\label{fig:Q21XX}
\end{figure*}

We next consider the reduced 21~cm auto-bispectrum as a function of ionization fraction, along with the 21~cm-$\delta$-$\delta$ cross-bispectrum.\footnote{The 21~cm-[CII]-[CII] cross-bispectrum is ultimately the quantity of interest, but as discussed in \S~\ref{ssec:CII_model}, we are unable to capture small mass halos and so we use the 21~cm-$\delta$-$\delta$ bispectrum as a proxy for
the 21~cm-[CII]-[CII] bispectrum. If the second-order predictions in \S~\ref{sec:approach} are correct, then we expect that all three of these bispectra differ only by a redshift-dependent constant.}
Our goals here are two-fold: first, we want to ascertain whether these quantities evolve strongly with $\avg{x_i}$, as expected if their evolution is driven, in part, by the  ``rise and fall'' behavior illustrated in Fig.~\ref{fig:ionization_history}. Second, we aim to see if the simulated bispectra follow the perturbative expectation that $Q_{21, \text{X}, \text{X}} = Q_{\delta, \delta, \delta}/\bto + C_{21, \text{X}, \text{X}}$.

Fig.~\ref{fig:Q21XX} shows that the various bispectra evolve strongly with ionization fraction. This is encouraging for our ultimate goal of using these measurements to extract information about the ionization history. The overall value and evolution of the reduced bispectra vary somewhat depending on which bispectrum is calculated, but this is expected since the values of $C_{21,21,21}$ and $C_{21,\delta,\delta}$ should differ by a factor of 3. Nevertheless, the overall shape of the two bispectra are similar, and the bispectra share the same ordering -- from smallest to largest $Q$ -- with increasing ionization fraction.

In order to test the perturbative formula, we determine the best fit $\bto$ and $C_{21,\text{X},\text{X}}$ for $Q_{21,\text{X},\text{X}}$. To carry out these fits we use the variance estimated from the scatter across our ten simulation realizations. We neglect the covariance between different wavevector bins, since our estimates of the off-diagonal elements of the covariance matrix are too noisy to be reliable. We expect this to have relatively little impact on the best fit parameters, but it does prevent us from assessing the overall goodness of fit of the perturbative formula since the effective number
of degrees of freedom are uncertain.
The inability to measure covariance also prevents us from accurately computing error bars on the inferred $\bto$ and $C_{21,\text{X},\text{X}}$. We defer this calculation to future studies.

\section{Checking the Perturbative Formulas}\label{sec:consistency_checks}

\begin{figure}
\includegraphics{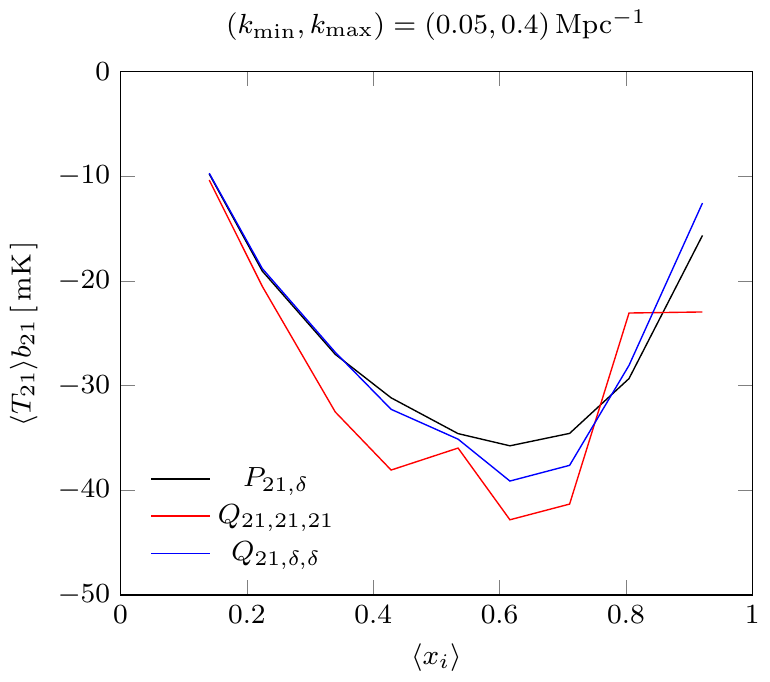}
\caption{Consistency between different estimates of $\bto$. The black line show the linear 21~cm bias factor, $\bto$, inferred from the cross-power spectrum, $\bto = P_{21, \delta}/P_{\delta, \delta}$ at $k = 0.05\,\Mpc^{-1}$, as a function of $\avg{x_i}$. These are compared with the linear bias inferred from $Q_{21,21,21}$ (red) and $Q_{21, \delta, \delta}$ (blue). 
}
\label{fig:b21_fromQ}
\end{figure}

An interesting test of the perturbative framework is to compare the values of $\bto$ and $C_{21,\text{X},\text{X}}$ inferred from a broad range of simulated two and three-point statistics. First, we compare the values of $\bto$ extracted from our fits to $Q_{21,\text{X},\text{X}}$ with those from the 21~cm-density cross-power spectrum. The results of this comparison are shown for a range of ionization fractions in Fig.~\ref{fig:b21_fromQ}. In general, all three approaches return similar values of $\bto$ and share the same qualitative evolution with ionization fraction. 
However, the $\bto$ estimates from the cross-power spectrum and the cross-bispectrum agree better with each other than with the auto-bispectrum inferences. This likely results because the 21~cm auto-bispectrum involves three significantly non-Gaussian fields (at least on scales smaller than the size of the ionized regions) and so our perturbative expansion is presumably less accurate for the auto-bispectrum. 
We find that by using $Q_{21,\delta,\delta}$, we are able to receover $\bto$ to within $5\%$ for $\avg{x_i} < 0.5$, and $10\%$ for $\avg{x_i} > 0.5$, except at the lowest redshift we tested ($z=6$). On the other hand, $Q_{21,21,21}$ is only able to recover $\bto$ to within $\sim 20\%$ accuracy for all of the EoR, although the precise number quoted here may be subject to noise in our
$Q_{21,21,21}$ estimates.

The agreement between the cross-power spectrum and cross-bispectrum estimates is best at $\avg{x_i} \lesssim 0.6$, at which point they separate a bit, before briefly coming together near $\avg{x_i} \sim 0.8$; they then split-off slightly again towards the tail-end of reionization.
The results in Fig.~\ref{fig:b21_fromQ} adopt triangles with wavevectors between
$k_{\tmin}=0.05\,\Mpc^{-1}$ and $k_{\tmax}=0.4\,\Mpc^{-1}$. In Appendix A (Fig.~\ref{fig:b21_results_app}) we investigate other choices for the range of wavevectors used in our fitting procedure. There we show that the values of $\bto$ inferred from $Q_{21,\delta,\delta}$ are insensitive to the precise choices of $k_{\tmin}$ and $k_{\tmax}$, while the results from $Q_{21,21,21}$ are noisier and show more sensitivity to the choice of scales.

\begin{figure}
\includegraphics{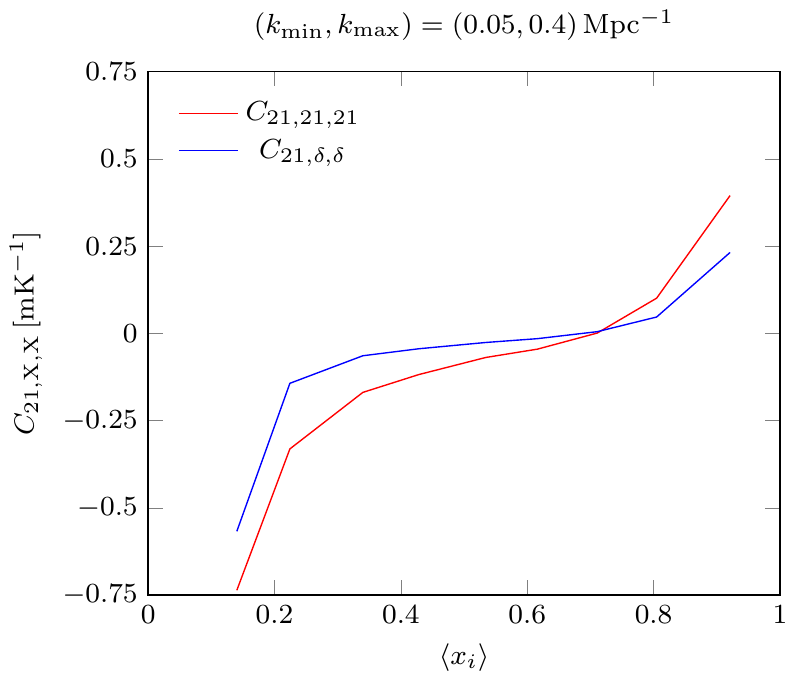}
\caption{
Evolution of the constant offset terms, $C_{21, X, X}$ with ionization fraction. 
Note that it appears that $C_{21,21,21} \sim C_{21,\delta,\delta}$ for $\avg{x_i} \sim 0.7$. This occurs, however, because both constant offset terms are very close to zero at this ionization fraction. See Fig.~\ref{fig:C21overC21del}.
}
\label{fig:offset}
\end{figure}

We can also test whether the constant offset terms agree with the formulas from perturbation theory (Eqs.~\ref{eq:C212121}-\ref{eq:C21CIICII}). Before considering this test, we note that the behavior of the constant offset term with redshift may offer an additional handle on the ionization history since e.g. $C_{21,21,21}$ is a measurable quantity.
Indeed, Fig.~\ref{fig:offset} shows that the constant offsets evolve fairly strongly with redshift/ionization fraction, and that the different offset terms evolve in qualitatively similar ways with average ionization fraction. However, note that $C_{21,\CII,\CII}$, which is most robust to foreground contamination, depends on
$b_{\CII}$ and $b_{\CII}^{(2)}$; that is, this quantity depends on both the [CII] and the 21~cm biasing and is therefore more challenging to interpret. 

\begin{figure}
\includegraphics{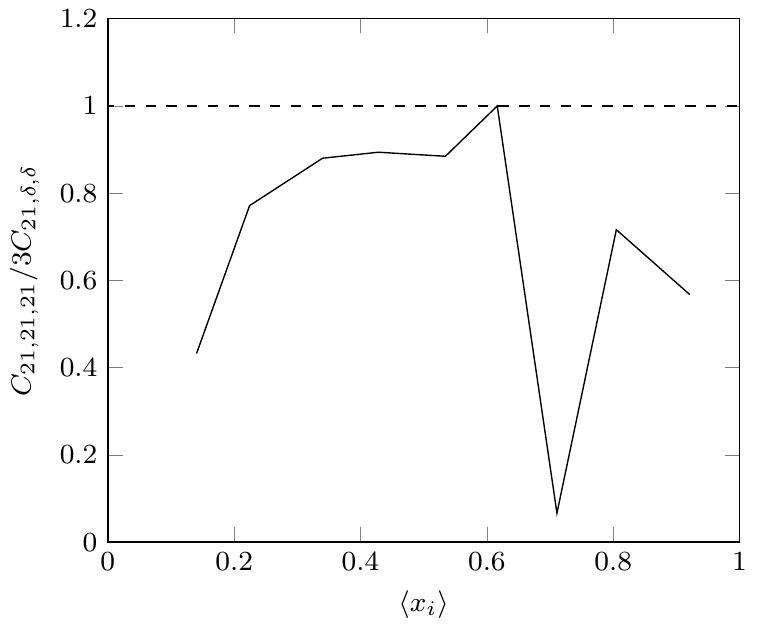}
\caption{
Consistency check on the constant offset terms. If second-order perturbation theory is accurate, we expect $C_{21, 21, 21}/3C_{21, \delta, \delta} = 1$. We find that this ratio is within $20\%$ of unity during the bulk of reionization. The dip at $\avg{x_i} \sim 0.7$ results because $C_{21,21,21}\text{,}C_{21,\delta,\delta} \sim 0$: at this stage of reionization, more simulation realizations would be required to accurately estimate
the ratio of these small values.
}
\label{fig:C21overC21del}
\end{figure}

The perturbative prediction (Eqs.~\ref{eq:C212121}~and~\ref{eq:C21deldel}) is that the constant offsets are related by  $C_{21,21,21} = 3 C_{21,\delta,\delta}$. Since we extracted each of these terms independently, we are able to test this statement without extracting $\bto^{(2)}$. We plot the ratio $C_{21,21,21}/3C_{21,\delta,\delta}$ in Fig.~\ref{fig:C21overC21del}. It is encouraging that for the much of the reionization process this ratio lies within $10-20\%$ of unity. The departures shown in the figure arise: i) at very early times (near $\avg{x_i} \sim 0.1$) when the correlation coefficient reverses sign (see \S \ref{ssec:21cmFAST}), ii) near $\avg{x_i} \sim 0.7$ when each of the constant offset terms happen to be close to zero and sample variance likely impacts our estimate of
this ratio, and iii) near the end of reionization when the ionized regions are large and the perturbative framework is expected to break down.

For completeness, we note that
with more detailed [CII] emission models, one could carry out further tests of the perturbative framework. Specifically, the perturbative formulas demand specific relationships between the constant offset terms for the 21~cm-[CII]-[CII] cross-bispectrum, the 21~cm-$\delta$-$\delta$ cross-bispectrum, the 21~cm auto-bispectrum, and the [CII] auto-bispectrum. Using Eqs.~\ref{eq:Q21XX}~-~\ref{eq:C21CIICII}, and analogous formulas for the [CII] intensity field, we have that,
\beq \label{eq:bCIIoverb21_x2dd}
\frac{3}{2}\frac{C_{21,\text{CII},\text{CII}}-C_{21,\delta,\delta}}{C_{\text{CII},\text{CII},\text{CII}}} = \frac{b_{\text{CII}}}{b_{21}}
\eeq
and, similarly,
\beq \label{eq:bCIIoverb21_a222}
\frac{3}{2}\frac{C_{21,\text{CII},\text{CII}}-\frac{1}{3}C_{21,21,21}}{C_{\text{CII},\text{CII},\text{CII}}} = \frac{b_{\text{CII}}}{b_{21}}\text{.}
\eeq
We comment in passing that another possibility is to extract $\avg{I_{\CII}}b_{\CII}$ from the analogous cross-bispectrum, $Q_{\CII,21,21}$. If line-intensity mapping experiments are performed using additional emission lines such as Ly-$\alpha$ or using rotational transitions from CO molecules, the cross-bispectra methodology may allow further inferences and consistency checks.

\section{Detectability}\label{sec:detectability}

We now discuss the prospects of measuring the 21~cm-[CII]-[CII] cross-bispectrum from future surveys. Here our aim is to obtain rough estimates of the joint sky coverage and noise power spectra
required for upcoming [CII] and 21~cm surveys to detect this signal. 

\subsection{[CII] Statistics}\label{ssec:cii_stats}

As discussed in \S~\ref{ssec:CII_model}, our current modeling assumes that the reduced cross-bispectrum $Q_{21,\CII,\CII}$ tracks $Q_{21,\delta,\delta}$ apart for an overall constant. 
In order to calculate the variance of $Q_{21,\CII,\CII}$, we require a model for the [CII] emission power spectrum and the cross-power spectrum with the 21~cm field. For this purpose, we assume a linear biasing model for the [CII] emission fluctuations:
\beq\label{eq:cii_psform}
\begin{split}
P_{\CII,\CII} &= \big(b_{\CII}\avg{I_{\CII}}\big)^2 P_{\delta,\delta} \\
P_{21,\CII} &= b_{\CII}\avg{I_{\CII}} P_{21,\delta}\text{.}
\end{split}
\eeq
Here $b_{\CII}$ is the luminosity-weighted bias of the halos that host [CII]-emitting galaxies. For simplicity, we suppose that $b_{\CII}=3$ at all redshifts of interest; our results are relatively insensitive to this choice. We follow \citet{lidz2016:remove_interloper} (which is based on \citet{Lidz11,Pullen:2013dir}) in calculating $\avg{I_{\CII}}$. Specifically, we assume a one-to-one mapping between star formation rate, SFR, and [CII] luminosity. The average specific intensity of the [CII] emission is:
\beq\label{eq:cii_intensity}
\avg{I_{\CII}}(z) = \frac{\epsilon_{\CII}}{4\pi \nu_{\text{rest}\text{,}\CII}} \frac{c}{H(z)}\text{,}
\eeq
where $\epsilon_{\CII}$ denotes the average co-moving emissivity (luminosity density) in the line, and $\nu_{\text{rest}\text{,}\CII}$ is the rest frame frequency of the [CII] line.
We have further assumed a delta function [CII] line emission profile. 

The one-to-one relation between [CII] luminosity and star-formation rate (SFR) we adopt is:  
\beq\label{eq:cii_halo_lum}
L = L_0 \frac{\text{SFR}}{1 M_{\sun}\text{yr}^{-1}}\text{,}
\eeq
with $L_0=6\times10^6 L_{\sun}$ \citep{righi08:coline,visbal10:3dclustering}. We neglect any redshift evolution in $L_0$ (see e.g. \citealt{lidz2016:remove_interloper} for a discussion).
We assume that the abundance of galaxies as a function of their SFR obeys a Schechter function form \citep{Schechter76}:
\beq \label{eq:schechter_sfr}
\phi(\SFR)\,\text{d}\SFR = \phi_* \left(\frac{\SFR}{\SFR_*}\right)^\alpha \exp\left[-\frac{\SFR}{\SFR_*}\right]\frac{\text{d}\SFR}{\SFR_*}\text{.}
\eeq
Here $\alpha$ denotes the faint-end slope, while $\phi_*$ and $\SFR_*$ are characteristic number densities and SFRs. The co-moving [CII] emissivity is then:
\beq\label{eq:com_emiss} 
\epsilon_{\CII} = \phi_* L_0 \frac{\SFR_*}{1\,M_{\sun}\text{yr}^{-1}}\Gamma(2+\alpha)\text{.}
\eeq
We use the Schechter SFR function parameters in \citet{Smit12}, determined using dust-corrected ultraviolet luminosity functions at redshifts of $z=2.2\text{, }3.8\text{, }5.0\text{, }5.9\text{, }6.8$. We find that the following functional form provides a reasonable approximation to the redshift evolution of the Schechter parameters:
\beq \label{eq:sfr_phi_fromsmit}
\begin{split}
\alpha &= -1.96 \\
\SFR_*(z) &= 134.39\,z^{-1.18}\,\si{\solarmass\per\year} \\
\phi_*(z) &= 0.00358\,z^{-0.85}\,\si{\per\mpc\cubed}\text{.}
\end{split}
\eeq
This procedure allows us to compute $\avg{I_{\CII}}$ at each relevant redshift. We find: $\avg{I_{\CII}} = 3.1 \times 10^2\,\Jystr\text{,}\, 7.1 \times 10^2\,\Jystr\text{,}\, 1.1 \times 10^3\,\Jystr$ for $z = 9.41\text{,}\, 7.37 \text{,}\, 6.43$, respectively.

\subsection{SNR Formula}\label{ssec:q_formula}

In order to estimate the signal-to-noise ratio (SNR), we require a formula for the cross-bispectrum variance, $\Var[Q_{21,\CII,\CII}]$. In calculating the variance, we suppose that the [CII] and 21~cm fluctuation fields obey Gaussian statistics. 
The variance of the cross-bispectrum (for $k_1 \neq k_2 \neq k_3$) is (see e.g. \citealt{Greig:2013Lya,joachimi2009:bisp_cov} for related calculations):
\beq \label{eq:var_bispectrum}
\begin{split}
&\Var[B_{21,\text{CII},\text{CII}} + \text{2 perm.}] = \\
& \frac{V_s}{N_t} \big(P_{\text{tot,21}}(k_1)P_{\text{tot,CII}}(k_2)P_{\text{tot,CII}}(k_3) + \text{2 perm.}\big)\text{,}
\end{split}
\eeq
where $P_\text{tot} = P + N$ is the total signal plus detector noise auto-power spectrum for each of the 21~cm and [CII] emission fields. The noise power spectrum for the 21~cm survey is subsequently denoted by $N_{21}$ and the [CII] noise power by $N_{\CII}$. We neglect contributions to the variance from residual foregrounds. 
Later we will add in the permutations on $k_1$, $k_2$, and $k_3$. Note that if any of $k_1$, $k_2$, or $k_3$ are equal, there are extra terms (which we write down explicitly in Appendix B). 
These extra terms are what give rise to the symmetry factor, $s_B$, in \citet{Greig:2013Lya} and related works.

Now, we use propagation of errors to recover $\Var[Q]$,
\beq \label{eq:var_q_prop}
\begin{split}
\Var[Q] &=  \left(\frac{\partial Q}{\partial B}\right)^2 \Var[B] \\
&+ \sum_{i=1}^{3} \left(\frac{\partial Q}{\partial P(k_i)}\right)^2 \Var[P(k_i)] \\
&+ \sum_{j=1}^{3} \frac{\partial Q}{\partial P(k_j)} \frac{\partial Q}{\partial B} \Cov[B,P(k_j)]\text{,}
\end{split}
\eeq
where the summations $i$ and $j$ span the three wavevector arguments, $k_1$, $k_2$, and $k_3$. 
Under the Gaussian approximation adopted here, $\Cov[B,P(k_i)]$ vanishes because it is a five-point function with vanishing unconnected pieces. Furthermore, we suppose that the term
involving $\Var[P(k_i)]$ is sub-dominant and neglect it in what follows. 
Finally, accounting for wavevector permutations, we arrive at our formula for the variance on $Q$:
\beq \label{eq:var_q}
\Var[Q] =  \frac{V_s}{N_t} \frac{\left(P_{\text{tot,21}}(k_1)P_{\text{tot,CII}}(k_2)P_{\text{tot,CII}}(k_3) + \text{2 perm.}\right)}{9\left(P_{21,\text{CII}}(k_1)P_{21,\text{CII}}(k_2) + \text{2 perm.}\right)^2}\text{,}
\eeq
where $V_s$ is the survey volume, and $N_t$ is the number of triangles, which we will derive shortly. Note that the $1/9$ factor comes from the $1/3$ in our definition of $Q_{21,\CII,\CII}$ in Eq.~\ref{eq:crossbisp_def}

In order to estimate the expected error bars on the parameters of interest, specifically $\bto$ and $C_{21,\CII,\CII}$, we use the Fisher matrix formalism. The Fisher matrix is calculated
as (e.g. \citealt{Greig:2013Lya}):
\beq \label{eq:fisher}
F_{ij} = \sum_{k_1\leq k_2\leq k_3} \frac{1}{\Var[Q](A, N_{21},N_{\CII})} \frac{\partial Q}{\partial \theta_{i}} \frac{\partial Q}{\partial \theta_{j}}\text{,}
\eeq
where $\theta_i$ denotes our two model parameters ($\bto$ and $C_{21,\CII,\CII}$.)

\subsection{Survey Parameters}\label{ssec:survey_param}

We adopt a simple, yet flexible, description for the upcoming 21~cm and [CII] surveys. First, we assume that the noise power spectrum for each survey is well-approximated by
isotropic, white-noise. In detail, this is an imperfect approximation, especially for the interferometric 21~cm observations in which case the noise power is a strong function of $k_\perp$; in reality, the 21~cm noise power is sensitive to the precise distribution of interferometric baselines. Our estimates here should be refined in future work. In our fiducial model, we match the amplitude of the white-noise power spectrum in each of the 21~cm and [CII] fields to their signal power at $k=0.05\,\Mpc^{-1}$, i.e. we set $N_{21}=P_{21,21}$ at this wavenumber, and similarly for $N_{\CII}$. This amounts to assuming that larger spatial scales are sample-variance dominated, while smaller-scales are limited by detector noise. For the survey volume, we assume that a common survey area, $A_\text{survey}$, is shared between the 21~cm and [CII] surveys, and adopt a redshift bin width of $\Delta z =0.3$. Our fiducial model takes $A_\text{survey}=50\,\text{deg}^2$. 

The number of triangles (with wavevectors within a given range) that fit into the overlap region of the two surveys may be calculated as:
\beq \label{eq:number_tri}
N_t = \frac{V_B}{V_{\text{fund}}^2}\text{.}
\eeq
Here $V_B$ is the 6D Fourier volume satisfying $\bm{k}_1+\bm{k}_2+\bm{k}_3 = \bm{0}$ of the 9D Fourier space spanned by $(\bm{k}_1\text{,}\bm{k}_2\text{,}\bm{k}_3)$. More specifically,
\beq \label{eq:vb_def}
V_B \equiv \int_{k_1} \text{d}^3q_1 \int_{k_2} \text{d}^3q_2 \int_{k_3} \text{d}^3q_3 \,\delta_{D}(\bm{q}_1 + \bm{q}_2 + \bm{q}_3)\text{,}
\eeq
where $\int_{k_i}$ refers to the integral over a wavevector bin centered around $|\bm{q_i}| = |\bm{k_i}|$. Assuming that $k_1 \geq k_2 \geq k_3$, one can show that (e.g. \citet{Greig:2013Lya}),
\beq \label{eq:vb_formula}
V_B \simeq 8 \pi^2 k_1 k_2 k_3 \,(\Delta k)^3 \,\theta(k_1, k_2, k_3)\text{,}
\eeq
where,
\beq \label{eq:symm_factor}
	\theta(k_1, k_2, k_3) =
	\begin{cases}
		\frac{1}{2} & \text{if $k_i = k_j + k_k$} \\
		1 & \text{if $k_i \neq k_j + k_k$,}
	\end{cases}
\eeq
is a symmetry factor ensuring that equivalent modes are not double counted. We adopt a bin size of $\Delta k = 0.03 \,\Mpc^{-1}$.
The quantity $V_\text{fund}$ in Eq.~\ref{eq:number_tri} denotes the $k$-space volume spanned by the fundamental modes of the joint survey regions. 
For this, we assume that the joint survey area is a square, giving:
\beq \label{eq:vfund}
V_\text{fund} = \frac{(2\pi)^3}{L_{\perp}^2L_{\parallel}}
\eeq
where $L_{\perp}$ and $L_{\parallel}$ are the co-moving extent of the survey volume perpendicular, parallel to the line of sight. We calculate $L_{\perp}$ at the central redshift of each redshift bin.

We then proceed to calculate the expected SNR on $\bto$, marginalizing over $C_{21,21,21}$, using the inverse Fisher matrix (Eq.~\ref{eq:fisher}). We perform these calculations at each of three different redshifts, while varying the survey parameters, $N_{\CII}$, $A_{\text{survey}}$, and $N_{21}$ around our fiducial values to explore how our forecasts depend on these inputs. 
 
 \begin{figure*}
\includegraphics{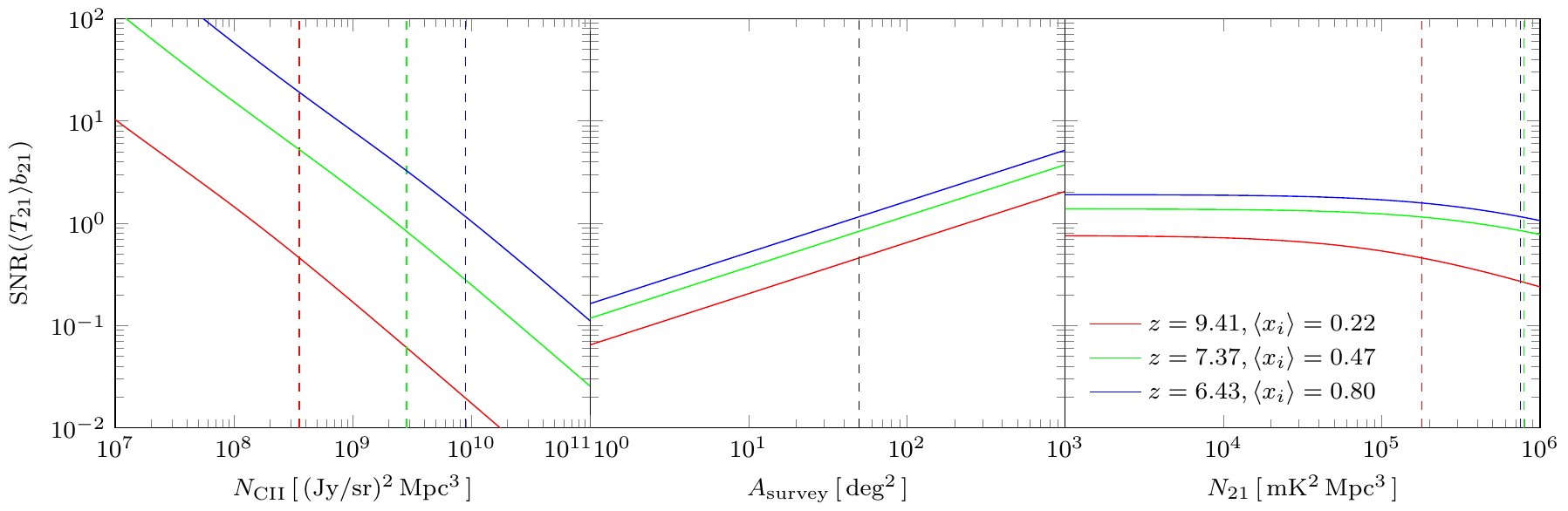}
\caption{Estimated SNR of $Q_{21,\text{CII},\text{CII}}$ as a function of ({\em Left-hand panel}) $N_\text{CII}$, ({\em Middle panel}) $A_\text{survey}$, and ({\em Right-hand panel}) $N_\text{21}$. When not varied, we assumed that $N_\text{CII} = P_{\CII,\CII}$ at $k = 0.05\,\Mpc^{-1}$, $A_\text{survey} = 50\,\text{deg}^2$ and $N_{21} = P_{21,21}$ at $k = 0.05\,\Mpc^{-1}$. Fiducial values are given by vertical dashed lines in each plot -- for $N_{21}$ and $N_{\CII}$ these are redshift dependent.}
\label{fig:SNR_estimate}
\end{figure*}
 
The results of these calculations are shown in Fig.~\ref{fig:SNR_estimate}. These estimates are generally encouraging: for example, in order to achieve a $10-\sigma$ detection (i.e. $\text{SNR}=10$) of the 21~cm-[CII]-[CII] bispectrum near the mid-point of reionization (at $z=7.37$ in this model) a $50\,\text{deg}^2$ survey with $N_{\CII} \lesssim 1.6 \times 10^8 \text\,(\text{Jy}/\text{sr})^2\,\Mpc^3$ is required.
This is more demanding than the noise expected for the ``Stage II'' [CII] survey described in \citet{silva15:prospects} which has a noise power spectrum of 
$N_{\CII} \sim 2.5 \times 10^9\,(\text{Jy}/\text{sr})^2\,\Mpc^3$ at $z=7$ \citep{lidz2016:remove_interloper}. Nonetheless, this improved sensitivity may be achievable with future improvements in detector technology, for example. As quantified in the figure, the requirements are somewhat less stringent at $z=6.43$ and somewhat more so at $z=9.41$. Note that our model assumes that the $L_{\CII}$-SFR relation is independent of redshift. This assumption is most suspect for this highest redshift bin, since the galaxies at this early time may have low metallicities and little [CII] emission. In any case, none of the currently planned [CII] emission surveys target such high redshifts \citep{kovetz2017:im_review}. 
The requirements on the 21~cm thermal noise appear less severe, with the SNR($\bto$) saturating for  $N_{21} \lesssim 10^5\,\mK^2\,\Mpc^3$. The results are less sensitive to the 21~cm noise power spectrum than the [CII] noise both because our cross-bispectrum involves only one 21~cm field (yet two [CII] fields) and since the 21~cm power spectrum becomes sample-variance limited at the wavenumbers of interest. The latter fact explains the saturation in the SNR with decreasing $N_{21}$. The required $N_{21}$ seems feasible since HERA-350 will {\em image} some large-scale
modes \citep{hera2017:sens}, implying that sample-variance limited sensitivity will be achieved on large scales.  

Another approach for improving the SNR($\bto$) is to increase the joint sky-coverage. For example, one possibility is to measure the 21~cm-Ly-$\alpha$-Ly-$\alpha$ bispectrum (rather than the 21~cm-[CII]-[CII] bispectrum considered here). This may be feasible with the planned all-sky survey SPHEREX \citep{spherex2018}, in which case one could match the entire $\sim 1440\, \text{deg}^2$ coverage of HERA-350 \citep{hera2017:sens}. The figure illustrates that our forecasts improve dramatically for wider survey areas. 

\begin{figure}
\includegraphics{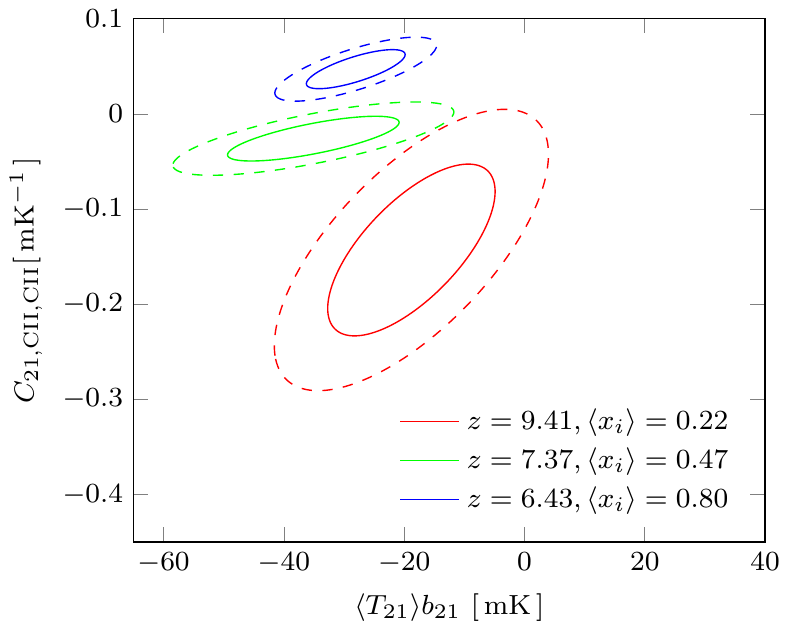}
\caption{Forecasted constraints in the $\bto$-$C_{21,\CII,\CII}$ plane at different redshifts. The solid/dashed ellipses show the expected $1-\sigma$/$2-\sigma$ contours at different redshifts, with each
contour centered on the values of $\bto$ and $C_{21,\CII,\CII}$ in our simulated models. These forecasts assume an ambitious, futuristic survey, with a sky coverage of $A = 1000\,\text{deg}^2$ and 
low detector noise with $N_{21}$ and $N_{\CII}$ matching $P_{21}$ and $P_{\CII}$ at the scale of $k=0.1\,\Mpc^{-1}$ (see text).
}
\label{fig:b21_C21_contours}
\end{figure}

To further explore the potential reach of our method, we consider more futuristic surveys and plot $1$ and $2-\sigma$ contours in the $\bto$-$C_{21,\CII,\CII}$ plane at different redshifts in
Fig.~\ref{fig:b21_C21_contours}. In this case, we set the joint survey area to be $A=1000\,\text{deg}^2$ and we match $N_{21}$ and $N_{\CII}$ to $P_{21}$ and $P_{\CII}$ at $k = 0.1\,\Mpc^{-1}$. This corresponds to noise values of $N_{21} = 1.7 \times 10^5\text{,}\,2.6 \times 10^5\text{,}\,7.6 \times 10^5\,\mK^2\,\Mpc^3$, and $N_{\CII} = 3.4 \times 10^9\text{,}\,1.1 \times 10^9\text{,}\,1.4 \times 10^8\,(\text{Jy}/\text{sr})^2\,\Mpc^3$ at $z=6.43\text{,}\,7.37\text{,}\,9.41$. This survey should indeed enable a sharp test of the ``rise and fall'' behavior from the cross-bispectrum. The contours also help illustrate the level of degeneracy between $\bto$ and $C_{21,\CII,\CII}$; these parameters are less covariant during the middle of reionization (when $\bto$ is large). This results because $Q_{21,\CII,\CII}$ has a flatter dependence on $\theta_{12}$ at this stage of reionization, which makes it easier to estimate $C_{21,\CII,\CII}$. 

\section{Conclusions}\label{sec:conclusion}

Our aim in this work has been to help circumvent two of the primary challenges associated with 21~cm fluctuation measurements of the EoR. The two concerns are that: residual foregrounds may produce errors in the inferred 21~cm auto-spectrum (or lead to a spurious detection), and that the 21~cm auto-spectrum is itself difficult to interpret given our imperfect models of the EoR. The first issue may, in part, be side-stepped by cross-correlating 21~cm fluctuation measurements with an additional tracer of the high redshift universe, such as line-intensity mapping data cubes in the [CII] emission line. The second concern is, in part, alleviated by identifying statistics that are amenable to an analytic treatment. 

Here, we proposed that the 21~cm-[CII]-[CII] cross-bispectrum may help in both regards. First, residual foregrounds only impact the variance of the cross-bispectrum and not the ensemble average (modulo foregrounds that are common to the two surveys.) Second, we showed that the configuration dependence of the 21~cm-$\delta$-$\delta$ cross-bispectrum (a proxy for the 21~cm-[CII]-[CII] cross-bispectrum which is more challenging to simulate) is fairly well described by second-order perturbation theory, especially provided one considers large spatial scales before the ionization fraction reaches $\avg{x_i} \sim 0.7$. Further, this may be used to extract the redshift
evolution of $\bto$ without resorting to reionization simulations. While such simulations are nevertheless required to understand the full implications of the $\bto$ measurements for the reionization history and the properties of the ionizing sources, we believe it is still valuable to consider upcoming 21~cm measurements in the context of cosmological perturbation theory. For example, the redshift evolution of $\bto$ inferred (on large scales) from the 21~cm auto-power spectrum, the 21~cm auto-bispectrum, and the 21~cm-[CII]-[CII] cross-bispectrum should each show the ``rise and fall'' signature. More generally, demonstrating the consistency of inferences from two and three-point statistics should help in establishing the robustness of initial 21~cm detections.
Unsurprisingly, the perturbative description is imperfect and future simulation efforts may help in exploiting information on small spatial scales, especially during the late stages of reionization. 

We made rough estimates of the survey specifications required to detect the 21~cm-[CII]-[CII] cross-bispectrum. Assuming a joint survey area of $A_\text{survey} = 50\, \text{deg}^2$, we found that a detection generally requires a more sensitive [CII] survey than the Stage-II [CII] emission survey described in \citet{silva15:prospects}. It would be interesting to refine our analysis by considering a more detailed treatment of the noise in each survey, by quantifying the impact of residual foregrounds on the cross-bispectrum variance, and by exploring to what extent foregrounds may be shared by the two surveys. Since the perturbative description is accurate on large spatial scales, this approach is most valuable if measurements can be made robustly at small wavenumbers ($k \lesssim 0.05-0.1\, \Mpc^{-1}$). 

In terms of modeling, we showed that the \texttt{L-PICOLA} code may be used in conjunction with 21cmFAST (replacing the Zel'dovich generated density field used in 21cmFAST by default) to improve the treatment of three-point statistics in EoR calculations, at only modest additional computational expense. In the future, it would be interesting to include a more detailed model for the [CII]-emission fluctuations: it is challenging to capture the large spatial scales of interest for our bispectrum calculations, while simultaneously resolving the small mass halos hosting [CII]-emitting galaxies. 
We considered only a single representative reionization model: we expect the perturbative description to work better if the ionized regions are smaller than in this model, while it will perform less well in scenarios with larger bubbles. 
While we considered only the 21~cm and [CII] emission lines in this work, the same methodology may be applied to other emission lines and it will be interesting to consider these prospects. 
In summary, higher-order statistics should help in extracting key information about the EoR from upcoming surveys.

\section*{Acknowledgements}
We thank the anonymous referee for providing helpful comments. We would like to thank Matthew McQuinn for helpful comments on a draft of this paper. AB would like to thank Congzhou M. Sha for helpful comments on code used in this work, and Todd Phillips for helpful discussions. AB was supported by the Roy \& Diana Vagelos Program in the Molecular Life Sciences and the Roy \& Diana Vagelos Challenge Award.

\bibliography{references}

\newpage

\section*{Appendix A: 21cm Bias Results}\label{sec:app_b21results}

\begin{figure}[h]
\includegraphics{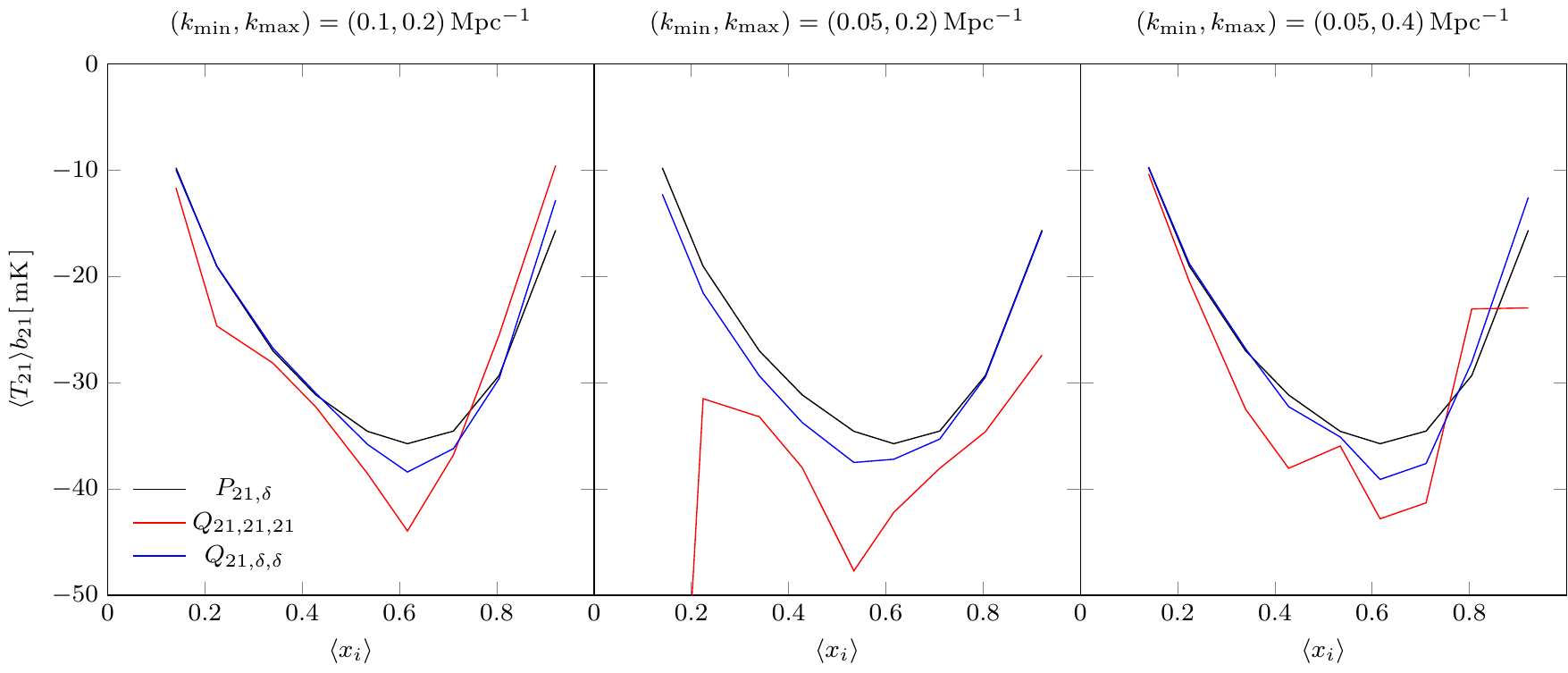}
\caption{Consistency between different estimates of $\bto$. The black curves show the linear 21~cm bias factor, $\bto$, inferred from the cross-power spectrum, $\bto = P_{21, \delta}/P_{\delta, \delta}$ on large scales, as a function of $\avg{x_i}$. These are compared with the linear bias inferred from $Q_{21,21,21}$ (red) and $Q_{21, \delta, \delta}$ (blue). {\em Left, Center:} Extracted $\bto$ for two different values of $k_{\tmin}$ and $k_{\tmax}$. {\em Right:} The plot from Fig.~\ref{fig:b21_fromQ} is reproduced here for convenience. Note that the values of $\bto$ inferred from the 21~cm auto-bispectrum at low $\avg{x_i}$ show a marked difference from the cross-power spectrum and cross-bispectrum inferences in the {\em Central panel}. We believe this relates to the early-phase transition in the sign of the 21 cm-density cross-correlation coefficient (see \S~\ref{ssec:21cmFAST})}

\label{fig:b21_results_app}
\end{figure}

Here we show how our $\bto$ results vary for different choices of $k_{\tmin}$ and $k_{\tmax}$ (Fig.~\ref{fig:b21_results_app}). As discussed in \S~\ref{ssec:bispectrum_algorithm}, the results are noisy
for small $k_{\tmin}$, while the perturbative description breaks down when $k_{\tmax}$ is too large. 

\section*{Appendix B: Cross-Bispectrum Variance}{\label{sec:app_bispvar}

Here we write down a formula for the cross-bispectrum variance in the case that some of the $k$-modes have equal magnitude. See e.g. \citet{joachimi2009:bisp_cov} for more details. Recall that we have enforced $k_1 \geq k_2 \geq k_3$.

\beq\label{varbisp_full}
	\Var[B_{21,\CII,\CII} + \text{2 perm.}]=
	\begin{cases}
		 P_{\text{tot,21}}(k_1)P_{\text{tot,CII}}(k_2)P_{\text{tot,CII}}(k_3) + \text{2 perm.} & \text{if $k_1 \neq k_2 \neq k_3$} \\
		 \\[12pt]
		 \begin{split}
		 &2 P_{\text{tot},21}(k_3)P_{\text{tot},\CII}(k_1)^2 \\
		 &+ 2P_{\text{tot},21}(k_1)P_{\text{tot},\CII}(k_3)P_{\text{tot},21}(k_1)\\
		 &+ 2P_{21,\CII}(k_1)^2P_{\text{tot},\CII}(k_3)
		 \end{split} & \text{if $k_1 = k_2 \neq k_3$} \\
		 \\[12pt]
		 \begin{split}
		 &2 P_{\text{tot},21}(k_1)P_{\text{tot},\CII}(k_2)^2 \\
		 &+ 2P_{\text{tot},21}(k_2)P_{\text{tot},\CII}(k_1)P_{\text{tot},21}(k_2)\\
		 &+ 2P_{21,\CII}(k_2)^2P_{\text{tot},\CII}(k_1)
		 \end{split} & \text{if $k_1 \neq k_2 = k_3$} \\
		 \\[12pt]
		 		 \begin{split}
		 &6P_{\text{tot,21}}(k_1)P_{\text{tot,CII}}(k_1)P_{\text{tot,CII}}(k_1) \\
		 & + 12 P_{21,\CII}(k_1)^2P_{\text{tot,CII}}(k_1)
		 \end{split} & \text{if $k_1 = k_2 = k_3$} \\
		 
	\end{cases}
\eeq

\end{document}